\documentclass[nofootinbib,aps,a4paper,letterpaper,superscriptaddress,
twocolumn,times,eqsecnum]{revtex4}
\usepackage{amsmath}
\usepackage{amsfonts}
\usepackage{booktabs}
\usepackage{multirow}
\usepackage{siunitx} 
\usepackage{adjustbox}
 \pdfoutput=1
\usepackage{graphicx}
\usepackage{color}
\usepackage{braket}
\usepackage{dcolumn}
\usepackage{bm,url}
\usepackage[linktocpage]{hyperref}
\usepackage{subfigure}
\usepackage{amsfonts}
\usepackage[usenames,dvipsnames,svgnames]{xcolor}  
\usepackage{hyperref}   
\definecolor{oxfordblue}{rgb}{0.0, 0.13, 0.28}
\definecolor{burgundy}{rgb}{0.5, 0.0, 0.13}
\definecolor{darkolivegreen}{rgb}{0.33, 0.42, 0.18}
\definecolor{darkblue}{rgb}{0,0,0.5}
\definecolor{richcarmine}{rgb}{0.84, 0.0, 0.25}
\definecolor{darkblue}{rgb}{0,0,0.5}
\definecolor{bluer}{rgb}{0.00,0.50,0.75}{}
\hypersetup{colorlinks=true, citecolor=red, linkcolor=blue,
 urlcolor = magenta, filecolor=magenta}

\newcommand{\mpl}{M_{\rm P}}

\newcommand{\lpl}{\ell_{\rm P}}
\begin{document}


\title{Polytropic wormholes}

\author{Remo Garattini}
\email{remo.garattini@unibg.it}
\affiliation{Universit\`a degli Studi di Bergamo, Dipartimento di Ingegneria e 
Scienze
Applicate, Viale Marconi 5, 24044 Dalmine (Bergamo) Italy and I.N.F.N.-
sezione di Milano, Milan, Italy.}

\author{Emmanuel N. Saridakis}
\email{msaridak@noa.gr}
\affiliation{Institute for Astronomy, Astrophysics, Space Applications and 
Remote Sensing, National Observatory of Athens, 15236 Penteli, Greece}
 \affiliation{Departamento de Matem\'{a}ticas, Universidad Cat\'{o}lica del 
  Norte, Avda. Angamos 0610, Casilla 1280, Antofagasta, Chile}
 \affiliation{CAS Key Laboratory for Research in Galaxies and Cosmology, 
School 
  of Astronomy and Space Science,
  University of Science and Technology of China, Hefei 230026, China}

\author{Athanasios G. Tzikas}
\email{athanasios.tzikas@unibg.it}
\affiliation{Universit\`a degli Studi di Bergamo, Dipartimento di Ingegneria e 
Scienze
Applicate, Viale Marconi 5, 24044 Dalmine (Bergamo) Italy.}

\begin{abstract}

Traversable wormholes in general relativity require non-standard matter sources,
making the identification of physically motivated equations of state 
particularly
important. We investigate wormholes supported by a polytropic equation of state,
considering homogeneous and inhomogeneous configurations within a unified
framework. We derive the corresponding solutions and analyze the effects of the
polytropic parameters on the geometry and energy conditions. In the homogeneous case, the polytropic construction yields a consistent wormhole interior whose geometry and matter content are governed by the constant polytropic parameters. For the
inhomogeneous case, we obtain a general analytical expression showing that the
geometry is completely determined by the radial polytropic coefficient
$\omega(r)$. For positive $\omega(r)$, the requirement for physically meaningful solutions naturally restricts the polytropic exponent to odd integer values. Using a power-law 
profile,
we construct explicit classes of solutions exhibiting distinct parameter
regimes and finite radial support. Interestingly enough, for an exponent 
$\alpha=2\gamma-3$, a
generalized   absurdly benign  traversable wormhole-like configuration emerges
naturally. Although the flare-out condition implies null-energy-condition
violation at the throat, the inhomogeneous framework allows its radial
distribution to be controlled. Our results establish a systematic connection
between polytropic matter and wormhole geometry, providing a flexible framework
for constructing compact wormholes with localized exotic matter.
 
\end{abstract}
\maketitle

\section{Introduction}
\label{sec:intro}

Traversable wormholes constitute one of the most interesting classes of 
solutions of the Einstein field equations, describing non-trivial spacetime 
geometries that connect two distinct regions of the Universe through a finite 
throat \cite{Morris:1988cz,Visser:1995cc,Morris:1988tu}. Beyond their 
mathematical interest, wormholes 
provide a theoretical laboratory for probing the interplay between geometry 
and matter, as well as for testing the limits of gravitational theories under 
extreme conditions. In particular, they are closely related to fundamental 
questions concerning causality, spacetime topology, and the possible existence 
of shortcuts through the spacetime manifold.

A central issue in wormhole physics is the nature of the matter required to 
support such configurations. Traversable wormholes 
typically require the violation of the null energy condition (NEC). This 
feature has motivated extensive research along two main directions. The first 
is to remain within general relativity and consider exotic or non-standard 
matter sources, such as anisotropic fluids, phantom energy, or more general 
effective equations of state, in an effort to minimize or control the required 
energy-condition violations \cite{Visser:1995cc,Morris:1988cz, Morris:1988tu, 
Visser:1989kh, Roman:1992xj, Armendariz-Picon:2002km,  
Lemos:2003jb, Kar:2004hc, Sushkov:2005kj, Lobo:2005us, Lobo:2005yv, 
Zaslavskii:2005fs, Lobo:2005vc, Rahaman:2005ur, Kuhfittig:2006rd, 
Chakraborty:2007na, Gonzalez:2009hn, Gonzalez:2008wd, Gonzalez:2008xk, 
Parsaei:2019hji, Garattini:2019ivd, Parsaei:2019utg,Alfaro:2024tdr}. The second 
direction is to 
consider extensions or modifications of gravity \cite{CANTATA:2021asi}, where 
the additional 
geometric degrees of freedom may effectively support wormhole geometries 
without the need for strongly exotic matter \cite{Lobo:2017cay,Lobo:2009ip, 
GarciaLobo:2010, GarciaLobo:2011, Lobo:2008zu, Lobo:2007qi, Harko:2013aya, 
SharifNawazish:2018, SamantaGodaniBamba:2020, GhoshMitra:2021, 
AgrawalMishra:2022, BaruahGoswami:2022, ZubairWaheedAhmad:2016, 
MoraesSahoo:2018a, SahooMoraesSahoo:2018, MoraesDePaulaCorrea:2019, 
MoraesSahoo:2019, BanerjeeJasimGhosh:2021, SahooMoraesLapola:2021, 
SaleemAslam:2023, Nashed:2026wdv, Sharif:2013exa, Kofinas:2015hla, 
MehdizadehZiaie:2017, Mehdizadeh:2017dhb, SaaidiTavakoli:2021, Landry:2025whg, 
ParsaeiRastgooSahoo:2022, MustafaHassanSahoo:2022, SokoliukHassanSahoo:2022, 
KiroriwalKumarMaurya:2023, RastgooParsaei:2024, HohmannKaranasou:2025, 
DehghaniDayyani:2009, Kanti:2011jz, MehdizadehZangenehLobo:2015, 
ZangenehLoboDehghani:2015, MehdizadehZiaie:2021, ChakrabortyChakraborty:2025, 
MunizEtAl:2025, Tsilioukas:2023tdw, Capozziello:2018mqy, Chatzifotis:2022mob, 
Barcelo:2000zf, KorolevSushkov:2014, BakopoulosCharmousisKanti:2022, 
BakopoulosChatzifotis:2023, Chatzifotis:2021hpg, CapozzielloHarkoKoivisto:2012, 
BambiCardenasOlmo:2016, RosaLemosLobo:2018, LoboOlmoOrazi:2020, 
KordZangenehLobo:2021, Rosa:2021, HarkoLoboMakSushkov:2015, Shaikh:2015, 
NandiIslamEvans:1997, ZiaieMehdizadeh:2024, Papantonopoulos:2019ugr}.

Within the first approach, a key ingredient is the choice of the equation of 
state  that characterizes the matter sector. In many studies, the 
supporting fluid is introduced in a phenomenological manner, often lacking a 
direct physical interpretation or exhibiting limited flexibility. This 
motivates the investigation of more structured and physically motivated 
equations of state. In this context, the polytropic equation of state, widely 
used in astrophysics for the description of stellar configurations and compact 
objects \cite{Tooper1964,Tooper1965}, provides a natural and versatile 
framework. Additionally, its ability to describe different physical regimes 
through a small number of parameters makes it particularly suitable for 
exploring non-trivial gravitational configurations.

Hence, traversable wormholes supported by fluids with a polytropic 
equation of state have attracted considerable attention in the literature. In 
particular, static and evolving wormhole solutions in four and higher 
dimensions have been constructed using polytropic matter \cite{Cataldo2013}, 
while combinations of barotropic and polytropic equations of state have been 
employed in order to reduce violations of the NEC \cite{VKD,Parsaeia2020}. 
Additionally, wormholes supported by polytropic phantom energy have been 
investigated in \cite{Jamil2009}, while thin-shell configurations with 
polytropic fluids have been shown to exhibit stability and asymptotically flat 
behavior \cite{ThinShell2024}. Moreover, earlier studies have also examined 
the 
properties of relativistic polytropes \cite{Tooper1964,Tooper1965}, as well as 
cosmological scenarios involving polytropic-type equations of state 
\cite{Lobo:2006mt}.

Despite this progress, several aspects of wormholes supported by polytropic 
matter remain insufficiently explored. In particular, a systematic analysis 
that treats both homogeneous and inhomogeneous polytropic equations of state 
within a unified framework, and examines the resulting classes of solutions 
together with their physical properties, is still lacking. Furthermore, the 
role of the polytropic parameters in determining the structure of the wormhole 
geometry, the behavior of the shape function, and the localization of 
energy-condition violations deserves a more detailed investigation.

In the present work we aim to address these issues by studying traversable 
wormhole solutions supported by a polytropic equation of state in a systematic 
and unified manner. We consider both homogeneous and inhomogeneous polytropic 
relations and derive the corresponding solutions of the gravitational field 
equations. Moreover, we analyze the resulting geometries by examining the 
throat conditions, the behavior of the shape function, and the constraints 
imposed by the flare-out condition. Particular emphasis is placed on the 
dependence of the solutions on the polytropic parameters and on the emergence 
of physically interesting subclasses, including configurations related to 
generalized absurdly benign traversable wormholes. In this way, the present 
analysis provides a unified characterization of how homogeneous and 
radially varying polytropic matter can support different classes of 
traversable wormhole geometries.

The plan of the work is as follows. In Sec.~\ref{Wormholegeometry}, we present 
the wormhole geometry and derive the corresponding field equations. In 
Sec.~\ref{Wormholepolytropic}, we introduce the polytropic equation of state 
and construct general wormhole solutions for both homogeneous and 
inhomogeneous cases. In Sec.~\ref{Classes}, we analyze various classes of 
solutions corresponding to the inhomogeneous polytropic parameter $\omega(r)$, 
constrain the physical parameters of the theory, and investigate their 
physical properties through the energy conditions and the behavior of the 
spacetime geometry. Finally, Sec.~\ref{Conclusions} is devoted to the 
conclusions. Throughout the paper, we use natural units in which 
$c=\hbar=1$ and $G=\lpl^2=\mpl^{-2}$.
 
 \section{Wormhole geometry and field equations}
\label{Wormholegeometry}

In this section we briefly review the basic framework of static and spherically
symmetric traversable wormholes and derive the corresponding gravitational
field equations. We additionally summarize the geometrical requirements that
a solution must satisfy in order to describe a traversable wormhole, as well
as their implications for the matter sector.

We consider the general Morris-Thorne metric describing a static and
spherically symmetric wormhole spacetime \cite{Morris:1988cz,Morris:1988tu}:
\begin{equation}
\mathrm{d}s^{2}=-e^{2\Phi(r)}\,\mathrm{d}t^{2}
+\frac{\mathrm{d}r^{2}}{1-b(r)/r}
+r^{2}\,(\mathrm{d}\theta^{2}+\sin^{2}{\theta}\,\mathrm{d}\varphi^{2})\,,
\label{metric}
\end{equation}
where $\Phi(r)$ is the redshift function associated with the gravitational
redshift, while $b(r)$ is the shape function, which determines the spatial
geometry of the wormhole. In particular, the latter controls the location and
geometry of the throat, whereas the behavior of $\Phi(r)$ determines whether
horizons are present. The radial coordinate $r$ is the areal radius and, in
each asymptotic region, it decreases down to its minimum value $r_0$ at the
wormhole throat and subsequently increases towards the other region.

In order to obtain traversable wormhole solutions, we consider the Einstein
field equations
\begin{equation}
G_{\mu\nu}=\kappa T_{\mu\nu}\,,
\label{EFE}
\end{equation}
where $\kappa=8\pi\ell_P^2$ and $T_{\mu\nu}$ is the stress-energy tensor.
For wormhole configurations, the matter content is naturally modeled as an
anisotropic fluid, with stress-energy tensor
\begin{equation}
T^{\mu}_{\ \nu}
=\mathrm{diag}\left[-\rho(r),p_r(r),p_t(r),p_t(r)\right]\,,
\end{equation}
where $\rho(r)$ is the energy density, $p_r(r)$ is the radial pressure, and
$p_t(r)$ is the tangential pressure of the fluid. The distinction between the
radial and tangential pressures is particularly relevant for wormhole
geometries, since the stress required to sustain the throat is in general
anisotropic.

Substituting the metric (\ref{metric}) into (\ref{EFE}), we obtain the
following system of differential equations:
\begin{eqnarray}
&&
\!\!\!\!\!\!\!\!\!\!\!\!\!\!\!\!\!\!\!
\frac{b'(r)}{r^{2}}=\kappa\rho(r)\,,
\label{rho0}
\\
&&\!\!\!\!\!\!\!\!\!\!\!\!\!\!\!\!\!\!\!
\frac{2}{r}\left(1-\frac{b(r)}{r}\right)\Phi'(r)
-\frac{b(r)}{r^{3}}
=\kappa p_r(r)\,,
\label{pr0}
\\ 
&&
\!\!\!\!\!\!\!\!\!\!\!\!\!\!\!\!\!\!\!\!\!
\left(1-\frac{b(r)}{r}\right)
\left[\Phi''(r)+\Phi'(r)\left(\Phi'(r)+\frac{1}{r}\right)\right]
\nonumber\\
&& \ \    -\frac{b'(r)r-b(r)}{2r^{2}}
\left(\Phi'(r)+\frac{1}{r}\right)
=\kappa p_t(r)\,.
\label{pt0}
\end{eqnarray}
Here and in the following, a prime denotes differentiation with respect to the
radial coordinate $r$. The above equations determine the relation between the
wormhole geometry and the matter quantities once an equation of state or an
additional condition is specified.

The field equations are complemented by conservation of the stress-energy
tensor, $\nabla_{\mu}T^{\mu}_{\ \nu}=0$, which yields
\begin{equation}
p_r'(r)=\frac{2}{r}\left[p_t(r)-p_r(r)\right]
-\left[\rho(r)+p_r(r)\right]\Phi'(r)\,.
\label{conservation}
\end{equation}
This relation may be interpreted as the hydrostatic-equilibrium equation for
the anisotropic fluid supporting the wormhole, with the first term on the
right-hand side encoding the effect of pressure anisotropy and the second one
the contribution associated with the redshift function.

Let us now summarize the geometrical conditions required for a physically
acceptable traversable wormhole. First, the existence of a throat at
$r=r_0$, corresponding to the minimum areal radius, requires
\begin{equation}
b(r_0)=r_0\,.
\label{throatcondition}
\end{equation}
Moreover, one requires
\begin{equation}
b(r)<r,\qquad r>r_0\,,
\label{signaturecondition}
\end{equation}
so that the radial metric component remains well defined on either side of
the throat. A crucial additional requirement is the flare-out condition,
which guarantees that the wormhole geometry opens outward away from the
throat. At $r=r_0$ this condition reduces to
\begin{equation}
b'(r_0)<1\,.
\label{flareout}
\end{equation}
Finally, traversability requires the absence of event horizons. Hence,
$\Phi(r)$ must remain finite throughout the wormhole spacetime, such that
$e^{2\Phi(r)}$ does not vanish. In the explicit constructions below we will
consider a constant redshift function, which automatically fulfills this
requirement.

An important consequence of these geometrical requirements concerns the
energy conditions. In fact, evaluating Eqs.~(\ref{rho0}) and (\ref{pr0}) at
the throat, and using $b(r_0)=r_0$, gives
\begin{equation}
\rho(r_0)+p_r(r_0)
=\frac{b'(r_0)-1}{\kappa r_0^2}\,.
\label{NECthroat}
\end{equation}
Therefore, the flare-out condition (\ref{flareout}) immediately implies
\begin{equation}
\rho(r_0)+p_r(r_0)<0\,,
\label{NECviolation}
\end{equation}
namely violation of the radial null energy condition at the throat. This is
the familiar connection between traversable wormhole geometry and exotic
matter within general relativity. Away from the throat, the extent and
magnitude of the energy-condition violation depend on the particular matter
configuration. The characterization of this behavior will be one of the main
aspects of the analysis that follows.

 \section{Wormholes with a polytropic equation of state}
\label{Wormholepolytropic}

In this section we investigate wormhole solutions supported by matter obeying a
polytropic equation of state. As discussed above, the properties of the matter
sector play a crucial role in determining the structure and physical viability
of traversable wormholes. In this context, the polytropic equation of state
provides a rich and physically motivated framework, widely used in
astrophysics and relativistic stellar structure.

We consider a polytropic equation of state relating the radial pressure to the
energy density in the form
\begin{equation}
p_r(r)=\omega \frac{\rho^\gamma(r)}{\rho_0^{\gamma-1}}\,,
\label{poly}
\end{equation}
where $\omega$ and $\gamma$ are dimensionless parameters. The constant
$\rho_0$, associated with a Planckian-like energy density
($\rho_0\sim\lpl^{-4}$), is introduced to ensure the correct dimensional
consistency of the physical quantities, while the energy density $\rho(r)$
retains dimensions of $L^{-4}$. The polytropic index $\gamma$ controls the
nonlinear dependence of the pressure on the energy density. In particular,
$\gamma>0$ corresponds to positive pressure for positive energy density
\cite{g0}, $\gamma=1$ describes an isothermal fluid \cite{g1}, while
$\gamma>1$ corresponds to standard polytropic configurations \cite{g2}.
Additionally, $\gamma\lesssim2$ is commonly associated with avoiding
superluminal sound speeds in relativistic polytropic configurations
\cite{g3}, whereas $\gamma<0$ is associated with Chaplygin-type fluids
\cite{g4,g5}. The parameter $\omega$ determines the overall amplitude and
sign of the radial pressure and thus controls the effective stiffness of the
polytropic fluid.

\subsection{Homogeneous polytropic equation}
\label{subHomogeneous}

We first consider the homogeneous polytropic case, in which the parameters
$\omega$ and $\gamma$ are constants. For simplicity, we focus on a constant
redshift function, which can be set to zero without loss of generality by a
rescaling of the time coordinate, namely $\Phi(r)=0$. Apart from simplifying
the field equations, this choice guarantees the absence of horizons associated
with the redshift function. Under this assumption, the Einstein field
equations (\ref{rho0})-(\ref{pt0}) reduce to
\begin{eqnarray}
\frac{b'(r)}{r^{2}} &=& \kappa \rho(r)\,,
\label{rho_h}\\
-\frac{b(r)}{r^{3}} &=& \kappa p_r(r)\,,
\label{pr_h}\\
-\frac{b'(r)r-b(r)}{2r^{3}} &=& \kappa p_t(r)\,,
\label{pt_h}
\end{eqnarray}
while the conservation equation simplifies to
\begin{equation}
p_r'(r)=\frac{2}{r}\left[p_t(r)-p_r(r)\right] .
\label{cons_h}
\end{equation}
Therefore, once the polytropic relation is imposed, the above system determines
the radial dependence of the matter variables and the corresponding wormhole
geometry.

Substituting the polytropic relation (\ref{poly}) into (\ref{pr_h}), we obtain
\begin{equation}
b(r)=-\kappa\omega r^3
\frac{\rho^\gamma(r)}{\rho_0^{\gamma-1}}\,,
\label{b0}
\end{equation}
which directly connects the shape function to the energy density.
Differentiating (\ref{b0}) and inserting into  
(\ref{rho_h}), we derive the   differential equation for the energy
density 
\begin{equation}
\rho'(r)=-\frac{1}{\gamma r}
\left[
3\rho(r)+
\frac{\rho^{2-\gamma}(r)}
{\omega\rho_0^{1-\gamma}}
\right]\,,
\label{rhodiff}
\end{equation}
which   admits the solution
\begin{equation}
\rho(r)=
\left[
C r^{\frac{3(1-\gamma)}{\gamma}}
-\frac{\rho_0^{\gamma-1}}{3\omega}
\right]^{\frac{1}{\gamma-1}} ,
\label{dens}
\end{equation}
where $C$ is an integration constant with dimensions
$L^{(\gamma-1)(3/\gamma-4)}$. Therefore, the radial profile of the energy
density is completely determined by the two polytropic parameters and a
single integration constant.

Substituting (\ref{dens}) into (\ref{b0}), we obtain the corresponding shape
function
\begin{equation}
b(r)=-\kappa\omega\rho_0^{1-\gamma}r^3
\left[
C r^{\frac{3(1-\gamma)}{\gamma}}
-\frac{\rho_0^{\gamma-1}}{3\omega}
\right]^{\frac{\gamma}{\gamma-1}}.
\label{bshape}
\end{equation}
We can now impose the geometrical conditions discussed in
Sec.~\ref{Wormholegeometry}. In particular, the throat condition
$b(r_0)=r_0$ fixes the integration constant according to
\begin{equation}
C=r_0^{\frac{3(\gamma-1)}{\gamma}}
\left[
\frac{\rho_0^{\gamma-1}}{3\omega}
-\left(
-\kappa\omega\rho_0^{1-\gamma}r_0^2
\right)^{\frac{1-\gamma}{\gamma}}
\right]\,.
\label{C}
\end{equation}
Inserting (\ref{C}) into (\ref{bshape}), the shape function can equivalently
be expressed as
\begin{equation}
b(r)=
\left[
r_0^{\frac{\gamma-1}{\gamma}}
+\frac{(-\kappa\rho_0)^{\frac{\gamma-1}{\gamma}}}
{3\omega^{1/\gamma}}
\left(
r_0^{\frac{3(\gamma-1)}{\gamma}}
-r^{\frac{3(\gamma-1)}{\gamma}}
\right)
\right]^{\frac{\gamma}{\gamma-1}} .
\label{bshape2}
\end{equation}
This form makes the dependence of the geometry on the throat radius and the
polytropic parameters explicit.

Furthermore, the flare-out condition $b'(r_0)<1$ leads to
\begin{equation}
(-1)^{1/\gamma}\kappa r_0^2
\left(
\frac{\rho_0^{\gamma-1}}
{\kappa\omega r_0^2}
\right)^{1/\gamma}<1\,.
\label{flarehom}
\end{equation}
For the branch satisfying $(-1)^{1/\gamma}=-1$ and $\omega>0$, the left-hand
side is negative and the flare-out condition is therefore automatically
satisfied. As will become clearer in the inhomogeneous construction, the
reality of these fractional powers places non-trivial restrictions on the
allowed values of the polytropic index.

On the other hand, the solution (\ref{bshape2}) is not asymptotically flat.
Indeed, the shape function does not exhibit the required large-distance
behavior and eventually  vanishes at a finite radius. It is therefore
natural to regard the solution as an interior wormhole geometry of finite
radial extent. Defining the cutoff radius $\bar r$ through $b(\bar r)=0$, one
obtains
\begin{eqnarray}
&&
\!\!\!\!
\bar{r}
=\left(3\omega 
C\rho_0^{1-\gamma}\right)^{\frac{\gamma}{3(\gamma-1)}}\nonumber\\
&&=r_0
\left[
1+
\frac{3\omega^{1/\gamma}}
{r_0^{(1+2\gamma)/\gamma}
(-\kappa\rho_0)^{(\gamma-1)/\gamma}}
\right]^{\frac{\gamma}{3(\gamma-1)}} . \ \
\label{rbar}
\end{eqnarray}
Thus, the wormhole interior is considered in the interval
$r_0\leq r\leq\bar r$ and may subsequently be matched to an asymptotically
flat exterior vacuum spacetime using the thin-shell formalism and the
corresponding junction conditions \cite{Isr}. A detailed construction of such
a matching lies beyond the scope of the present work.

Finally, the tangential pressure follows from (\ref{pt_h}) and reads
\begin{equation}
p_t(r)=-\frac{1}{2}
\left[
\omega\frac{\rho^\gamma(r)}{\rho_0^{\gamma-1}}
+\rho(r)
\right]\,,
\label{pthom}
\end{equation}
where $\rho(r)$ is given by (\ref{dens}). Hence, despite imposing the
polytropic equation of state only in the radial direction, the tangential
pressure is fixed consistently by the gravitational field equations. As a
final consistency check, direct substitution of the above expressions into
(\ref{cons_h}) shows that the conservation equation is identically satisfied.
Therefore, the homogeneous polytropic construction provides a self-consistent
wormhole interior, whose geometrical and matter properties are controlled by
the polytropic parameters and the throat scale.

\subsection{Inhomogeneous polytropic equation}
\label{subInhomogeneous}

We now proceed to the more general case in which the polytropic coefficient
$\omega$ is allowed to vary radially. In particular, instead
of the homogeneous relation (\ref{poly}), we consider the inhomogeneous
polytropic equation of state
\begin{equation}
p_r(r)=\omega(r)\frac{\rho^\gamma(r)}{\rho_0^{\gamma-1}}\,,
\label{poly_inh}
\end{equation}
where the constant parameter $\omega$ is promoted to a function $\omega(r)$
of the radial coordinate. This generalization introduces an additional
functional freedom in the matter sector, allowing the effective stiffness of
the supporting fluid to vary throughout the wormhole geometry.

Substituting (\ref{poly_inh}) into the radial Einstein equation
(\ref{pr0}), we obtain
\begin{equation}
\Phi'(r)=\frac{\kappa r^3 p_r(r)+b(r)}
{2r\left[r-b(r)\right]}
=
\frac{\kappa r^3\omega(r)\rho^\gamma(r)/\rho_0^{\gamma-1}+b(r)}
{2r\left[r-b(r)\right]}\,.
\label{phiinh}
\end{equation}
As in the homogeneous case, we focus on the subclass with constant redshift
function. The constant can be set to zero by a rescaling of the time
coordinate, and hence we take $\Phi(r)=0$. Imposing $\Phi'(r)=0$ in
(\ref{phiinh}) leads to
\begin{equation}
\kappa\,\omega(r)\frac{\rho^\gamma(r)}{\rho_0^{\gamma-1}}
+\frac{b(r)}{r^3}=0\,.
\label{condPhi0}
\end{equation}
Using the first Einstein equation (\ref{rho0}), namely
$\rho(r)=b'(r)/(\kappa r^2)$, relation (\ref{condPhi0}) can be rewritten as
a differential equation involving only the shape function:
\begin{equation}
\kappa\frac{\omega(r)}{\rho_0^{\gamma-1}}
\left(\frac{b'(r)}{\kappa r^2}\right)^\gamma
+\frac{b(r)}{r^3}=0\,.
\label{bde}
\end{equation}
This equation constitutes the basic differential relation governing the
inhomogeneous polytropic wormhole configurations. In contrast to the
homogeneous case, the radial dependence of $\omega(r)$ directly determines
the radial behavior of the resulting geometry.

Solving (\ref{bde}) for the shape function and imposing the throat condition
$b(r_0)=r_0$, we obtain the general solution
{\small{
\begin{equation}
b(r)=\!\left[
r_0^{\frac{\gamma-1}{\gamma}}
+(-1)^{\frac{1}{\gamma}}
\left(\frac{\gamma-1}{\gamma}\right)
(\kappa\rho_0)^{\frac{\gamma\!-\!1}{\gamma}}\!
\int_{r_0}^{r}
\frac{\tilde r^{\frac{2\gamma-3}{\gamma}}}
{\omega(\tilde r)^{1/\gamma}}\,
\mathrm{d}\tilde r
\right]^{\frac{\gamma}{\gamma-1}}.
\label{bp}
\end{equation}}}
Therefore, once the function $\omega(r)$ is specified, the corresponding
wormhole geometry is completely determined through (\ref{bp}). This is one
of the useful features of the inhomogeneous construction, since different
radial profiles of the polytropic coefficient naturally generate different
classes of wormhole geometries.

Differentiating (\ref{bp}), we find
{\small{
\begin{eqnarray}
&&
\!\!\!\!\!\!\!\!\!\!\!\!\!\!\!\!\!\!\!\!
b'(r)=\frac{\gamma}{\gamma-1}
\left[
(-1)^{\frac{1}{\gamma}}
\left(\frac{\gamma-1}{\gamma}\right)
(\kappa\rho_0)^{\frac{\gamma-1}{\gamma}}
\frac{r^{\frac{2\gamma-3}{\gamma}}}
{\omega(r)^{1/\gamma}}
\right]
\nonumber\\
&& \!\!\!\!\!\!\!\!\!\!\!\!\!\!\!\!\!\!\!\!
\cdot
\left[
r_0^{\frac{\gamma\!-\!1}{\gamma}}
\!+\!(-1)^{\frac{1}{\gamma}}
\left(\frac{\gamma\!-\!1}{\gamma}\right)\!
(\kappa\rho_0)^{\frac{\gamma-1}{\gamma}}\!
\int_{r_0}^{r}
\frac{\tilde r^{\frac{2\gamma-3}{\gamma}}}
{\omega(\tilde r)^{1/\gamma}}\,
\mathrm{d}\tilde r
\right]^{\frac{1}{\gamma-1}}\!\!.
\label{bpprime}
\end{eqnarray}}}
Evaluating this expression at the throat, and using $b(r_0)=r_0$, yields
\begin{equation}
b'(r_0)=
(-1)^{\frac{1}{\gamma}}
(\kappa\rho_0)^{\frac{\gamma-1}{\gamma}}
\left(\frac{r_0^{\frac{2(\gamma-1)}{\gamma}}}
{\omega(r_0)^{1/\gamma}}\right)\,.
\label{Flare}
\end{equation}
Hence, the flare-out condition $b'(r_0)<1$ imposes direct constraints on
the allowed values of $\gamma$, $\rho_0$, $r_0$, and, in particular, on the
sign and magnitude of $\omega(r_0)$. Notice that the combination entering
(\ref{Flare}) is dimensionless, as required.

At this stage it is useful to distinguish between odd and even integer values
of the polytropic index. For odd $\gamma$, namely
\begin{equation}
\gamma=2n+1\,,
\end{equation}
relation (\ref{Flare}) becomes
\begin{equation}
b'(r_0)=-
(\kappa\rho_0)^{\frac{2n}{2n+1}}
\frac{r_0^{\frac{4n}{2n+1}}}
{\omega(r_0)^{1/(2n+1)}}\,.
\label{FlareOdd}
\end{equation}
Therefore, for $\omega(r_0)>0$ one has $b'(r_0)<0$, and the flare-out
condition is automatically satisfied. On the other hand, for
$\omega(r_0)<0$, the derivative becomes positive and the flare-out
condition leads to the non-trivial bound
\begin{equation}
(\kappa\rho_0)^{\frac{2n}{2n+1}}
\frac{r_0^{\frac{4n}{2n+1}}}
{|\omega(r_0)|^{1/(2n+1)}}<1\,.
\label{FlareOdd2}
\end{equation}

For even values, $\gamma=2n$, the reality of the solution imposes stronger
restrictions. In particular, the combination entering the $\gamma$-th root
must be non-negative, which in the present construction requires
$\omega(r_0)<0$. Thus, positive $\omega(r)$ naturally selects the odd-integer
branch of the polytropic exponent. Since the detailed behavior of the geometry
depends on the full radial profile of $\omega(r)$, explicit realizations will
be investigated in Sec.~\ref{Classes}.

In summary, the inhomogeneous polytropic framework provides a broad class of
wormhole solutions characterized by the radial profile of $\omega(r)$. The
general solution (\ref{bp}) makes explicit the direct correspondence between
the choice of the equation-of-state function and the resulting shape function,
while the throat relation (\ref{Flare}) determines the basic restrictions
required by the flare-out condition. This framework therefore provides a
systematic way of constructing and classifying inhomogeneous polytropic
wormhole geometries.

 \section{Classes of solutions and physical properties}
\label{Classes}

In the previous section we derived the general solution for wormhole geometries
supported by an inhomogeneous polytropic equation of state. We now proceed to
construct explicit classes of solutions by specifying the radial profile of
the equation-of-state function $\omega(r)$. Our aim is to identify
representative choices for which the resulting geometry can be obtained
analytically and its physical properties can be examined explicitly.

\subsection{Specific choices of $\omega(r)$}

A simple and physically transparent choice is to consider a power-law profile
for the inhomogeneous polytropic coefficient, namely
\begin{equation}
\omega(r)=\left(\frac{r}{r_0}\right)^\alpha,
\label{o(r)}
\end{equation}
where $\alpha$ is a real parameter controlling the radial evolution of the
equation of state. This ansatz preserves the positivity of $\omega(r)$ for
$r>r_0$, while introducing a controlled radial dependence in the matter
sector. In particular, $\alpha>0$ corresponds to an increasing polytropic
coefficient away from the throat, whereas $\alpha<0$ describes a decreasing
one.

Substituting (\ref{o(r)}) into the general solution (\ref{bp}), we obtain
\begin{eqnarray}
&&
\!\!\!\!\!\!\!\!\!\!\!\!\!\!\!\!\!\!\!\!\!
b(r)=r_0\bigg[
1+(-1)^{1/\gamma}\left(\frac{\gamma-1}{\gamma}\right)
r_0^{\alpha/\gamma}\nonumber\\
&& \ \ \ \cdot
\left(\frac{\kappa\rho_0}{r_0}\right)^{\frac{\gamma-1}{\gamma}}
\int_{r_0}^{r}
\tilde r^{\frac{2\gamma-3-\alpha}{\gamma}}\,
\mathrm{d}\tilde r
\bigg]^{\frac{\gamma}{\gamma-1}}.
\label{bpower}
\end{eqnarray}
Hence, the exponent $\alpha$ directly modifies the radial dependence of the
integrand and, consequently, controls the qualitative behavior of the shape
function.

As discussed in Sec.~\ref{Wormholepolytropic}, in order to obtain real
solutions for positive $\omega(r)$ we focus on odd integer values of the
polytropic index, $
\gamma=2n+1$, with $ n\in\mathbb{N}$.
In this case, expression (\ref{bpower}) becomes
\begin{eqnarray}
&&
\!\!\!\!\!\!\!\!\!\!\!\!\!\!\!\!\!\!\!\!\!b(r)=r_0\bigg[
1-\left(\frac{2n}{2n+1}\right)
r_0^{\frac{\alpha}{2n+1}}\nonumber\\
&& \ \ \ \cdot
\left(\frac{\kappa\rho_0}{r_0}\right)^{\frac{2n}{2n+1}}
\int_{r_0}^{r}
\tilde r^{\frac{4n-1-\alpha}{2n+1}}\,
\mathrm{d}\tilde r
\bigg]^{\frac{2n+1}{2n}}.
\label{bp1}
\end{eqnarray}

A particularly interesting benchmark case arises for
\begin{equation}
\alpha=2\gamma-3=4n-1\,,
\label{alphabench}
\end{equation}
for which the power of $\tilde r$ in the integrand vanishes. The integration
can then be performed immediately, giving
\begin{equation}\!
b(r)=r_0\left[
1-\left(\frac{2n}{2n\!+\!1}\right)
\frac{r_0\kappa\rho_0}
{(r_0^2\kappa\rho_0)^{1/(2n+1)}}(r\!-\!r_0)
\right]^{\frac{2n+1}{2n}}\!.
\label{bp2}
\end{equation}
Defining
\begin{equation}
\mu=\left(\frac{2n}{2n+1}\right)
\frac{r_0\kappa\rho_0}
{(r_0^2\kappa\rho_0)^{1/(2n+1)}}
\qquad \mathrm{with}\qquad n\geq1\,,
\label{mu}
\end{equation}
the shape function takes the compact form
\begin{equation}
b(r)=r_0\left[1-\mu(r-r_0)\right]^{\frac{2n+1}{2n}}.
\label{GABTWform}
\end{equation}
In particular, its derivative at the throat is
\begin{equation}
b'(r_0)
=-\frac{2n+1}{2n}\,\mu r_0<0\,,
\label{flareGABTW}
\end{equation}
and therefore the flare-out condition is automatically satisfied for
$\mu>0$.

Expression (\ref{GABTWform}) has the form of a generalized absurdly benign
traversable wormhole (GABTW) of type I \cite{GABTW}, provided that the
solution is supplemented by
\begin{equation}
b(r)=0
\qquad \mathrm{for}\qquad
r\geq\bar r=r_0+\frac{1}{\mu}\,.
\label{cutGABTW}
\end{equation}
Thus, the non-trivial wormhole geometry has compact radial support,
$r_0\leq r\leq\bar r$, while the spacetime outside this region is flat for
the constant-redshift choice adopted here.

\begin{figure}[ht]
 \includegraphics[scale=0.43]{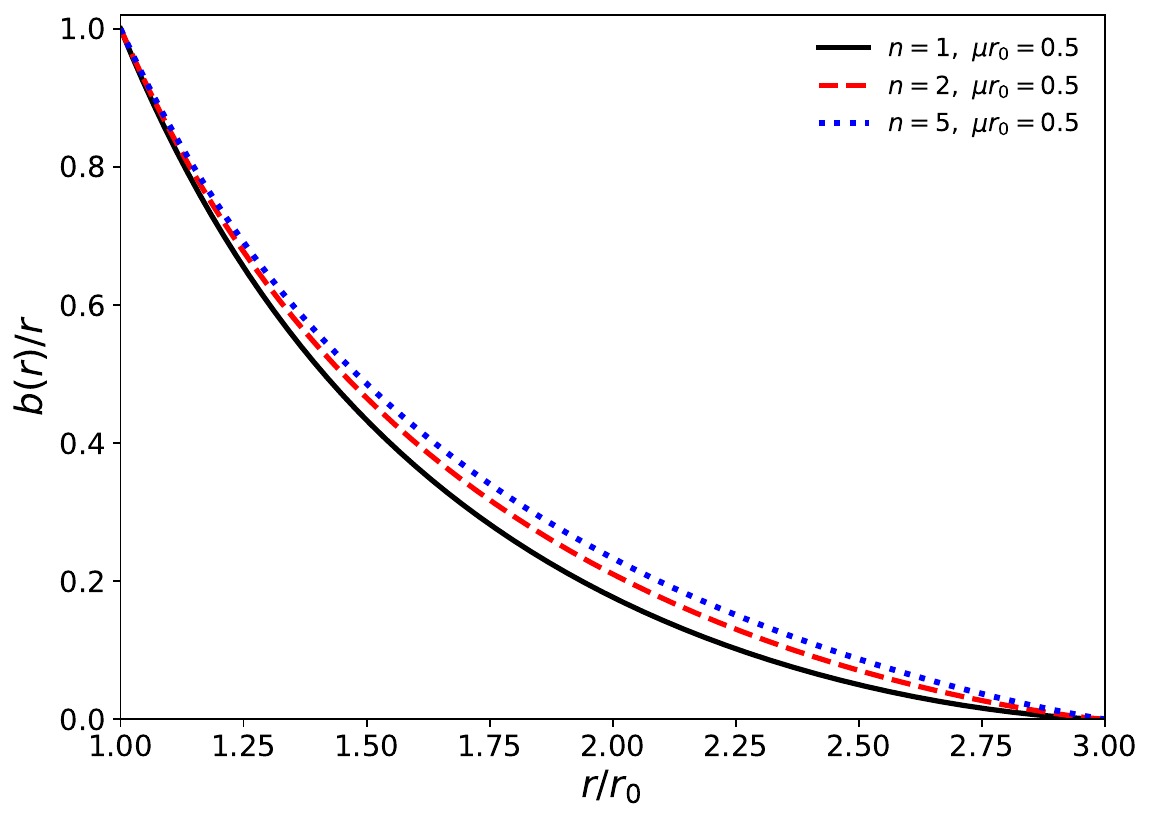}
\caption{{\it
Normalized shape function $b(r)/r$ for the GABTW-like solution
(\ref{GABTWform}), for $\mu r_0=0.5$ and three representative values of the
polytropic parameter: $n=1$ (black solid), $n=2$ (red dashed), and $n=5$
(blue dotted). In all cases, $b(r)/r$ decreases monotonically from unity at
the throat and vanishes at the finite boundary $\bar r/r_0=3$. The different
curves illustrate the dependence of the wormhole geometry on the polytropic
index.}}
\label{fig1}
\end{figure}

 The behavior of the corresponding shape function is illustrated in
Fig. \ref{fig1}. In particular, we depict the normalized quantity
$b(r)/r$ for $\mu r_0=0.5$ and three representative values of the
polytropic parameter, namely $n=1,2,$ and $5$. In all cases, $b(r)/r$
decreases monotonically from unity at the throat and reaches zero at the
finite boundary $\bar r/r_0=3$, explicitly displaying the compact radial
support of the geometry. Moreover, the comparison between the different
values of $n$ shows that the polytropic index controls the radial profile
of the wormhole geometry, while leaving the location of the external
boundary unchanged for fixed $\mu r_0$.

Notice also that both $b(r)$ and
$b'(r)$ vanish as $r\rightarrow\bar r$, so that the matter quantities obtained
from the field equations continuously approach zero at the boundary. Further
properties of this configuration are investigated in
Appendix~\ref{App}.

It is interesting to examine also the limiting behavior of this class. In the
limit $n\rightarrow\infty$,  expression (\ref{GABTWform}) reduces to
\begin{equation}
b_\infty(r)
=r_0\left[1-r_0\kappa\rho_0(r-r_0)\right],
\label{binfty}
\end{equation}
which is linear in the radial coordinate and no longer belongs to the GABTW
class. Hence, the power-law ansatz (\ref{o(r)}) generates a family of
wormhole geometries whose structure depends non-trivially on the polytropic
index and includes, for the particular choice
$\alpha=2\gamma-3$, a compact-support GABTW configuration.

Finally, we mention that for the benchmark case (\ref{alphabench}) the
power-law profile $\omega(r)$ grows without bound if formally extended to
$r\rightarrow\infty$. This large-distance behavior, however, is not relevant
for the wormhole interior, since the physical solution terminates at the
finite radius $\bar r$ given in (\ref{cutGABTW}). In the following subsection
we investigate how the geometry changes when $\alpha$ departs from the
benchmark value (\ref{alphabench}).

\subsection{Parameter regimes}

We now examine the resulting solutions for different ranges of the exponent
$\alpha$. Since the power of the radial coordinate appearing in the integrand
of (\ref{bp1}) depends explicitly on $\alpha$, qualitatively different
geometrical behaviors can arise. We therefore distinguish the three regimes
$\alpha>2\gamma-3$, $0<\alpha<2\gamma-3$, and $\alpha<0$.

\subsubsection{Case $\alpha>2\gamma-3$}

Let us first consider
\begin{equation}
\alpha>2\gamma-3=4n-1\,.
\end{equation}
It is convenient to parameterize $
\alpha=4n-1+x$ with $ x>0$.
Then (\ref{bp1}) takes the form
\begin{eqnarray}
&&
\!\!\!\!\!\!\!\!\!\!\!\!\!\!\!\!\!\!\!\!\!
b(r)=r_0\bigg[
1-\left(\frac{2n}{2n+1}\right)
r_0^{\frac{4n-1+x}{2n+1}}\nonumber\\
&& \ \ \ \cdot
\left(\frac{\kappa\rho_0}{r_0}\right)^{\frac{2n}{2n+1}}
\int_{r_0}^{r}
\tilde r^{-\frac{x}{2n+1}}\,\mathrm{d}\tilde r
\bigg]^{\frac{2n+1}{2n}}.
\label{breg1}
\end{eqnarray}
The qualitative behavior is therefore determined by the value of $x$, which
controls whether the integral produces a power-law or logarithmic dependence.

A useful subclass is obtained by setting
\begin{equation}
x=k(2n+1),
\qquad k>0\,,
\end{equation}
for which two cases can be distinguished.

\paragraph{Subcase $k=1$:}

For $k=1$, the integral becomes logarithmic and one obtains
{\small{
\begin{equation}
b(r)=r_0\left[
1-\left(\frac{2n}{2n+1}\right)
r_0^{\frac{6n}{2n+1}}
\left(\frac{\kappa\rho_0}{r_0}\right)^{\frac{2n}{2n+1}}
\ln\left(\frac{r}{r_0}\right)
\right]^{\frac{2n+1}{2n}}.
\label{blog}
\end{equation}}}
Moreover, the corresponding matter variables are
{\small{
\begin{align}
&\rho(r)\!=\!-\frac{r_0^{1+\frac{3}{\alpha}}}{\kappa r^3}
\left(\frac{\kappa\rho_0}{r_0}\right)^{1/\alpha}\!
\left[
1\!-\!\frac{r_0^{3/\alpha}}{\alpha}
\left(\frac{\kappa\rho_0}{r_0}\right)^{1/\alpha}
\ln\left(\frac{r}{r_0}\right)
\right]^{\alpha\!-\!1},
\\
&p_r(r)\!=\!-\frac{r_0}{\kappa r^3}
\left[
1-\frac{r_0^{3/\alpha}}{\alpha}
\left(\frac{\kappa\rho_0}{r_0}\right)^{1/\alpha}
\ln\left(\frac{r}{r_0}\right)
\right]^\alpha,
\\
&p_t(r)\!=\!-\frac{\rho(r)+p_r(r)}{2}\,.
\end{align}}}
In this case the solution possesses a finite radial domain. In particular,
the shape function vanishes at
\begin{equation}
\bar r=r_0\exp\left[
\left(\frac{2n+1}{2n}\right)
r_0^{-\frac{6n}{2n+1}}
\left(\frac{\kappa\rho_0}{r_0}\right)^{-\frac{2n}{2n+1}}
\right],
\label{rbar1}
\end{equation}
and therefore the physically relevant wormhole interior is restricted to
$r_0<r<\bar r$. Beyond this radius the continuation of the above expression
gives $b(r)<0$ and does not belong to the wormhole configuration considered
here.

For the generic case $x\neq2n+1$, integration of (\ref{breg1}) gives
\begin{eqnarray}
&&
\!\!\!\!\!\!\!\!\!\!\!\!\!\!\!\!\!\!\!\!\!
b(r)=r_0\Bigg[
1-
(2n) 
r_0^{\frac{4n-1+x}{2n+1}}
\left(\frac{\kappa\rho_0}{r_0}\right)^{\frac{2n}{2n+1}}\nonumber\\
&& \ \ \ \cdot\left(
\frac{
r^{\frac{2n+1-x}{2n+1}}
-r_0^{\frac{2n+1-x}{2n+1}}
}{
2n+1-x
}\right)
\Bigg]^{\frac{2n+1}{2n}}.
\label{breg1gen}
\end{eqnarray}
The corresponding $\rho(r)$, $p_r(r)$, and $p_t(r)$ follow directly from the
field equations. Depending on the parameters, the physically relevant branch
is again determined by requiring the shape function to remain real and to
satisfy the wormhole conditions discussed in Sec.~\ref{Wormholegeometry}.

\paragraph{Subcase $k\neq1$:}

For $k\neq1$, expression (\ref{breg1}) becomes
\begin{eqnarray}
&&
\!\!\!\!\!\!\!\!\!\!\!\!\!\!\!\!\!\!\!\!\!
b(r) 
=r_0\bigg[
1-\left(\frac{2n}{2n+1}\right)
(\kappa\rho_0)^{\frac{2n}{2n+1}}\nonumber\\
&& \ \ \ \cdot 
r_0^{\frac{(2n-1)(1+k)}{2n+1}}
\left(
\frac{1}{r^{k-1}}-\frac{1}{r_0^{k-1}}
\right)
\bigg]^{\frac{2n+1}{2n}}.
\label{bkneq1}
\end{eqnarray}
Hence, departures from the logarithmic case lead to a power-law dependence
of the shape function on the radial coordinate. The allowed radial domain is
again fixed by imposing reality of $b(r)$ together with $b(r)<r$ outside the
throat.

\subsubsection{ Case $0<\alpha<2\gamma-3$}

We now consider
\begin{equation}
0<\alpha<2\gamma-3=4n-1\,.
\end{equation}
It is convenient to write
$
\alpha=4n-1-y$ with $0<y<4n-1$.
Then (\ref{bp1}) becomes
\begin{eqnarray}
&&
\!\!\!\!\!\!\!\!\!\!\!\!\!\!\!\!\!\!\!\!\!
b(r)=r_0\bigg[
1-\left(\frac{2n}{2n+1}\right)
r_0^{\frac{4n-1-y}{2n+1}}\nonumber\\
&& \ \ \ \cdot 
\left(\frac{\kappa\rho_0}{r_0}\right)^{\frac{2n}{2n+1}}
\int_{r_0}^{r}
\tilde r^{\frac{y}{2n+1}}\,\mathrm{d}\tilde r
\bigg]^{\frac{2n+1}{2n}}.
\label{breg2}
\end{eqnarray}
In contrast to the previous regime, the power of $\tilde r$ in the integrand
is now positive. Thus, the radial contribution to the shape function grows
with increasing distance from the throat.

For the special choice
\begin{equation}
y=s(2n+1),
\end{equation}
we again distinguish two subcases.

\paragraph{Subcase $s=1$:}

For $s=1$, expression (\ref{breg2}) becomes
{\small{
\begin{equation}
b(r)=r_0\left[
1-\left(\frac{2n}{2n\!+\!1}\right)
r_0^{\frac{2n-2}{2n+1}}
\left(\frac{\kappa\rho_0}{r_0}\right)^{\frac{2n}{2n+1}}
\left(\frac{r^2\!-\!r_0^2}{2}\right)
\right]^{\frac{2n+1}{2n}},
\label{breg2special}
\end{equation}}}
with matter variables
\begin{align}
\rho(r)=&-\frac{1}{\kappa r}
r_0^{\frac{-1+4n}{1+2n}}
\left(\frac{\kappa\rho_0}{r_0}\right)^{\frac{2n}{1+2n}}\nonumber\\
& \cdot
\left[
1-\frac{
n\,r_0^{\frac{2n-2}{1+2n}}(r^2\!-\!r_0^2)
}{1+2n}
\left(\frac{\kappa\rho_0}{r_0}\right)^{\frac{2n}{1+2n}}
\right]^{\frac{1}{2n}},
\\
p_r(r)=&-\frac{r_0}{\kappa r^3}\!
\left[
1-\frac{
n\,r_0^{\frac{2n-2}{1+2n}}(r^2\!-\!r_0^2)
}{1+2n}
\left(\frac{\kappa\rho_0}{r_0}\right)^{\frac{2n}{1\!+\!2n}}
\right]^{1+\frac{1}{2n}},
\\
p_t(r)=&-\frac{\rho(r)+p_r(r)}{2}\,,
\end{align}
which are valid for $n>1$. The explicit expressions show again that the
matter profile is fully fixed once the geometrical parameters are specified.

\paragraph{Subcase $s\neq1$:}

For $s\neq1$, expression (\ref{breg2}) for a generic value of $y$ becomes
\begin{eqnarray}
&&
\!\!\!\!\!\!\!\!\!\!\!\!\!\!\!\!\!\!\!\!\!
b(r)=r_0\bigg[
1-\left(\frac{2n}{2n+1+y}\right)
r_0^{\frac{2n-2}{2n+1}}
\left(\frac{\kappa\rho_0}{r_0}\right)^{\frac{2n}{2n+1}}\nonumber\\
&& \ \ \   \cdot
\left(
r^{\frac{2n+1+y}{2n+1}}
-r_0^{\frac{2n+1+y}{2n+1}}
\right)
\bigg]^{\frac{2n+1}{2n}}.
\label{breg2gen}
\end{eqnarray}
For the branches considered here the geometry typically terminates at a
finite radius $\bar r$, determined by $b(\bar r)=0$, and the resulting
wormhole solution is therefore naturally interpreted as a finite interior
configuration.

\subsubsection{Case $\alpha<0$}

Finally, we consider negative values of the exponent,
\begin{equation}
\alpha<0\,.
\end{equation}
In this case  (\ref{bp1}) becomes
\begin{eqnarray}
&&
\!\!\!\!\!\!\!\!\!\!\!\!\!\!\!\!\!\!\!\!\!
b(r)=r_0\bigg[
1-\left(\frac{2n}{2n+1}\right)
r_0^{-\frac{\alpha}{2n+1}}\nonumber\\
&& \ \ \   \cdot
\left(\frac{\kappa\rho_0}{r_0}\right)^{\frac{2n}{2n+1}}
\int_{r_0}^{r}
\tilde r^{\frac{4n-1+\alpha}{2n+1}}\,\mathrm{d}\tilde r
\bigg]^{\frac{2n+1}{2n}}.
\label{breg3}
\end{eqnarray}
This regime differs from the previous ones because the polytropic coefficient
$\omega(r)$ decreases away from the throat. The resulting matter profile may
therefore become progressively weaker with increasing $r$, although the
actual behavior of the stress-energy components depends on the simultaneous
radial evolution of $\rho(r)$. The physically admissible domain must be
determined by requiring the shape function to remain real and to satisfy
$b(r)<r$ for $r>r_0$.

In summary, the power-law ansatz (\ref{o(r)}) gives rise to several distinct
classes of wormhole geometries, whose behavior is controlled by the exponent
$\alpha$. For a broad range of the parameter space, the physically admissible
solutions possess a finite radial extent, with the shape function reaching
zero at a finite radius. This provides a natural mechanism for confining the
non-trivial wormhole geometry to a bounded region.

Before concluding this subsection, it is instructive to consider separately a
non-integer value of the polytropic exponent directly from the general
inhomogeneous solution (\ref{bp}). In particular, for
\begin{equation}
\gamma=\frac{3}{2},
\end{equation}
one obtains
\begin{equation}
b(r)=\left[
r_0^{1/3}
+\frac{(\kappa\rho_0)^{1/3}}{3}
\int_{r_0}^{r}
\frac{\mathrm{d}\tilde r}{\omega(\tilde r)^{2/3}}
\right]^3.
\label{bgamma32}
\end{equation}
This example lies outside the odd-integer branch considered above and is
included in order to illustrate more generally the interplay between the
radial behavior of $\omega(r)$ and the asymptotic properties of the shape
function.

For instance, choosing
\begin{equation}
\omega(r)=
\left(\frac{r}{r+r_0}\right)^{3/2}
\label{omega32a}
\end{equation}
gives
\begin{align}
b(r)
=&\left[
r_0^{1/3}
+\frac{(\kappa\rho_0)^{1/3}}{3}
\left(
r-r_0+r_0\ln\frac{r}{r_0}
\right)
\right]^3.
\label{bgamma32a}
\end{align}
Clearly, this shape function grows at large $r$ and therefore does not satisfy
the asymptotic-flatness requirement $b(r)/r\rightarrow0$.

On the other hand, adopting
\begin{equation}
\omega(r)=
\left[
\frac{r^3}{r_0^2(r+r_0)}
\right]^{3/2},
\label{omega32b}
\end{equation}
one finds
\begin{align}
b(r)
=&\left[
r_0^{1/3}
+\frac{(\kappa\rho_0)^{1/3}}{3}
\left(
\frac{3r_0}{2}
-\frac{r_0^2}{r}
-\frac{r_0^3}{2r^2}
\right)
\right]^3.
\label{bgamma32b}
\end{align}
In this case $b(r)$ approaches a finite constant as $r\rightarrow\infty$,
and therefore $b(r)/r\rightarrow0$, yielding an asymptotically flat geometry.
However, the price to pay is that $\omega(r)$ diverges at large radii.

These examples illustrate an apparent tension between obtaining a well-behaved
radial polytropic profile and recovering asymptotic flatness for this
particular value of $\gamma$. While the examples above do not constitute a
general no-go theorem, they show that satisfying both requirements
simultaneously is non-trivial and strongly constrains the admissible form of
$\omega(r)$.

\subsection{Physical properties and energy conditions}

We now proceed to examine the main physical features of the wormhole solutions
derived above, focusing on their domain of validity, the energy conditions, and
the role of the polytropic parameters in determining the resulting geometry.

A common feature of a broad class of the homogeneous and inhomogeneous
polytropic solutions obtained above is the existence of a finite radial domain
in which the wormhole interior is defined. In particular, for the branches
considered in the previous subsections the shape function typically vanishes at
a finite radius $r=\bar r$, beyond which the continuation of the corresponding
solution gives $b(r)<0$. We therefore restrict the wormhole interior to the
interval
\begin{equation}
r_0\leq r\leq \bar r\,,
\end{equation}
and regard $\bar r$ as an external boundary of the non-trivial geometry.
The resulting interior configuration can in principle be matched to an exterior
vacuum spacetime by means of the appropriate junction conditions
\cite{Isr}. Hence, these solutions naturally describe wormhole geometries of
finite radial extent, supported by a localized matter distribution.

\begin{figure}[!]
 \includegraphics[scale=0.44]{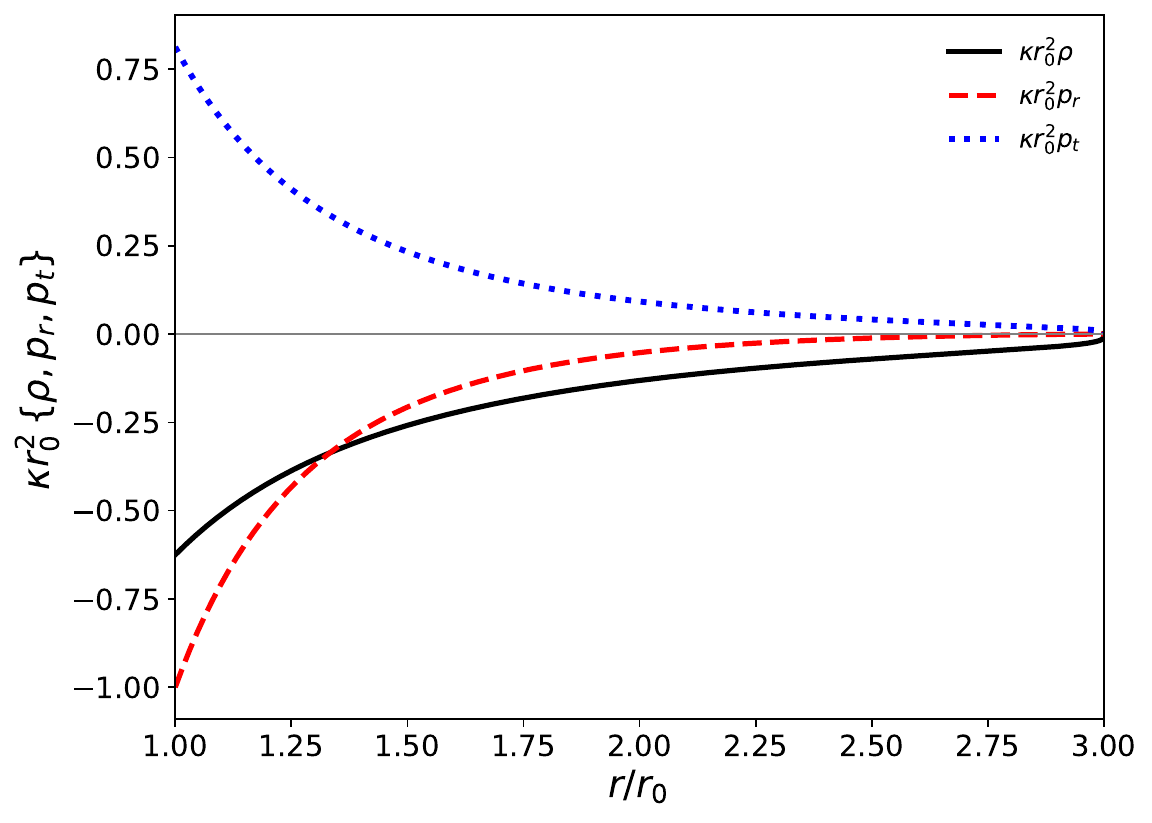}
\caption{{\it
Radial profiles of the dimensionless matter components
$\kappa r_0^2\rho$ (black solid), $\kappa r_0^2p_r$ (red dashed), and
$\kappa r_0^2p_t$ (blue dotted) for the GABTW-like solution
(\ref{GABTWform}), for the representative choice $n=2$ and
$\mu r_0=0.5$. The distinct behavior of the radial and tangential
pressures illustrates the anisotropic nature of the supporting matter.
All components approach zero at the finite boundary
$\bar r/r_0=3$, consistently with the compact support of the
matter distribution.
 }}
\label{fig2}
\end{figure}

This behavior is illustrated explicitly in Fig.~\ref{fig2}, where we
show the dimensionless matter components $\kappa r_0^2\rho$,
$\kappa r_0^2p_r$, and $\kappa r_0^2p_t$ for the GABTW-like solution
(\ref{GABTWform}), choosing the representative values $n=2$ and
$\mu r_0=0.5$. The different radial profiles of $p_r$ and $p_t$
clearly display the anisotropic character of the matter supporting the
wormhole. Moreover, all three components continuously approach zero at
the finite boundary $\bar r/r_0=3$, explicitly confirming the
localization of the matter distribution within the compact wormhole
interior.

Let us next examine the energy conditions. As discussed in
Sec.~\ref{Wormholegeometry}, within general relativity the flare-out condition
immediately implies violation of the radial null energy condition at the
throat. In particular,
\begin{equation}
\left[\rho(r)+p_r(r)\right]_{r=r_0}
=
\frac{b'(r_0)-1}{\kappa r_0^2}<0\,.
\label{NECthroatIV}
\end{equation}
Thus, exotic matter is necessarily required at least in the vicinity of the
throat for the solutions studied here.

For the homogeneous polytropic solutions, the radial NEC combination takes
the form
\begin{equation}
\rho(r)+p_r(r)
=
\rho(r)
\left[
1+\omega
\frac{\rho^{\gamma-1}(r)}{\rho_0^{\gamma-1}}
\right].
\label{NEChom}
\end{equation}
Its radial behavior is therefore controlled by both the polytropic parameters
and the corresponding density profile (\ref{dens}). Although
Eq.~(\ref{NECthroatIV}) guarantees that this quantity is negative at the
throat, its behavior away from $r_0$ depends on the specific values of the
parameters and on the radial evolution of $\rho(r)$.

Similarly, in the inhomogeneous case one obtains
\begin{equation}
\rho(r)+p_r(r)
=
\rho(r)
\left[
1+
\frac{\omega(r)}{\rho_0^{\gamma-1}}
\rho^{\gamma-1}(r)
\right].
\label{NECinh}
\end{equation}
Hence, in addition to the density profile, the amount and radial extent of the
NEC violation are directly affected by the functional form of $\omega(r)$.
This additional freedom provides a mechanism for controlling the distribution
of the exotic matter supporting the wormhole. In particular, suitable
radially varying profiles may reduce the region over which the NEC is violated,
although the precise extent must be determined for each specific solution.

For the zero-redshift configurations considered throughout the explicit
analysis, the tangential pressure satisfies
\begin{equation}
p_t(r)=-\frac{1}{2}\left[\rho(r)+p_r(r)\right],
\label{ptNEC}
\end{equation}
and consequently the tangential NEC combination becomes
\begin{equation}
\rho(r)+p_t(r)
=
\frac{1}{2}\left[\rho(r)-p_r(r)\right].
\label{NECtang}
\end{equation}
Thus, the radial and tangential null energy conditions need not exhibit the
same behavior, reflecting the intrinsically anisotropic character of the
matter source.

\begin{figure}[!]
 \includegraphics[scale=0.44]{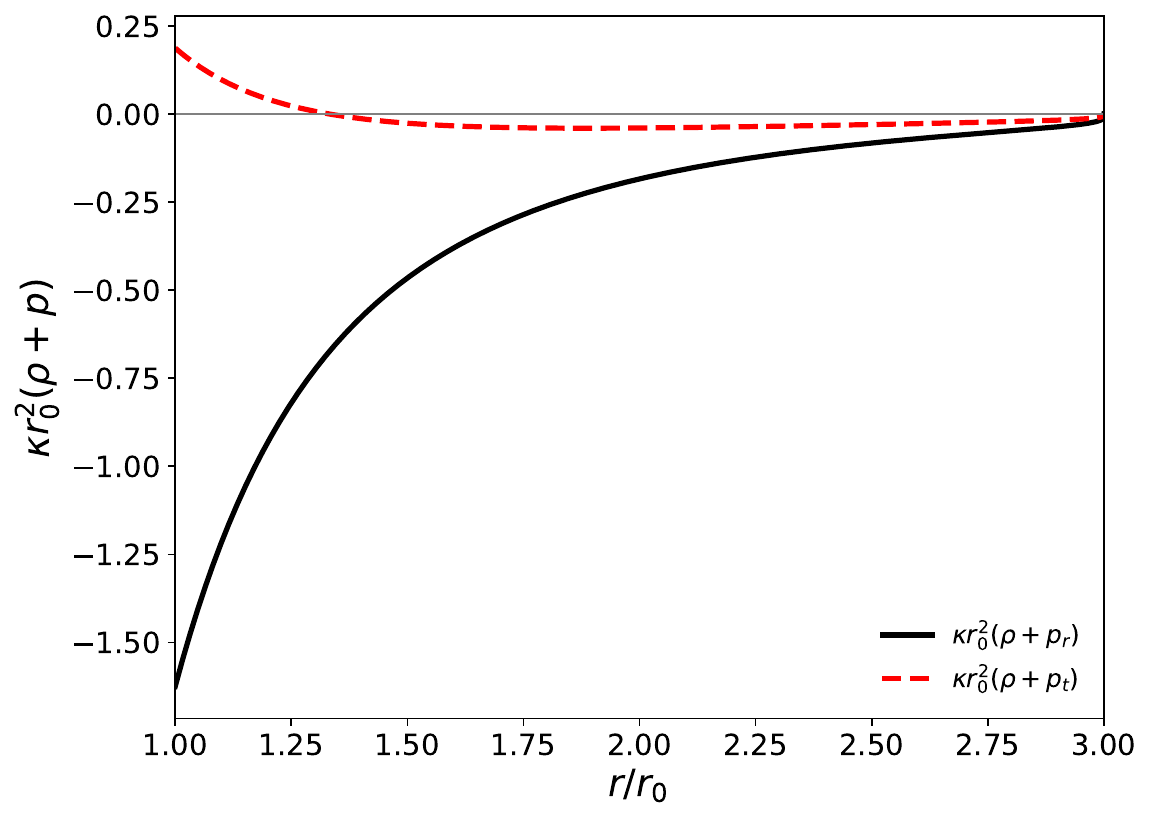}
\caption{\textit{Radial profiles of the dimensionless null-energy-condition
combinations $\kappa r_0^2(\rho+p_r)$ (black solid) and
$\kappa r_0^2(\rho+p_t)$ (red dashed) for the GABTW-like solution
(\ref{GABTWform}), for the representative choice $n=2$ and
$\mu r_0=0.5$. The radial NEC is violated throughout the wormhole
interior and approaches zero at the finite boundary, while the
tangential NEC is satisfied near the throat and exhibits only a mild
violation at intermediate radii. Both combinations approach zero at
$r/r_0=3$, consistently with the compact support of the matter
distribution.}}
\label{fig3}
\end{figure}

The behavior of the two NEC combinations is illustrated in
Fig.~\ref{fig3} for the GABTW-like solution (\ref{GABTWform}),
again choosing $n=2$ and $\mu r_0=0.5$. As expected from the
flare-out condition, the radial NEC is violated at the throat and
remains violated throughout the wormhole interior, approaching zero
at the finite boundary. In contrast, the tangential NEC is satisfied
near the throat and exhibits only a mild violation at intermediate
radii. Both combinations continuously approach zero at
$\bar r/r_0=3$, consistently with the compact support of the matter
distribution.

An important aspect of the present framework is the role of the polytropic
parameters in shaping the wormhole geometry. The index $\gamma$ controls the
nonlinear response of the pressure to variations in the energy density, while
$\omega$ determines the overall strength of this relation in the homogeneous
case. In the inhomogeneous construction, the function $\omega(r)$ introduces
an additional radial dependence, and for the power-law ansatz
(\ref{o(r)}) the exponent $\alpha$ controls whether the effective polytropic
coefficient increases or decreases away from the throat. As demonstrated in
the previous subsection, different choices of these parameters lead to
qualitatively distinct shape functions and radial domains.


In summary, the polytropic framework considered in this work gives rise to
wormhole solutions whose geometry and matter content can be controlled through
the polytropic parameters. The flare-out condition necessarily enforces radial
NEC violation at the throat, while the homogeneous and especially the
inhomogeneous constructions allow considerable freedom in determining how this
violation evolves away from it. Together with the finite radial extent found
for a broad class of solutions, these properties illustrate the flexibility of
polytropic matter as a source of traversable wormhole geometries.

\section{Conclusions}
\label{Conclusions}
 
Traversable wormholes constitute particularly interesting non-trivial
spacetime configurations, both from the geometrical point of view and as a
framework for investigating the properties of matter under extreme
gravitational conditions. Within general relativity, their construction is
closely connected to the nature of the matter source, since the flare-out
condition necessarily implies violation of the radial null energy condition
at the throat. It is therefore important to investigate whether sufficiently
structured and physically motivated matter sectors can provide a systematic
framework for supporting wormhole geometries and, at the same time, allow
control over their geometrical and physical properties. Motivated by this
question, in this work we have investigated traversable wormholes supported by
a polytropic equation of state.

In particular, we considered both homogeneous and inhomogeneous polytropic
matter within a unified framework. In the homogeneous case, the radial
pressure is determined by constant polytropic parameters, and we obtained
analytically the corresponding energy-density profile and shape function. By
imposing the throat and flare-out conditions, we determined the restrictions
on the integration constants and model parameters. The resulting solutions
generically describe wormhole interiors of finite radial extent rather than
globally asymptotically flat geometries, and therefore they can naturally be
regarded as interior configurations to be matched to an exterior vacuum
spacetime.

We then generalized the construction by promoting the polytropic coefficient
to a function of the radial coordinate. This inhomogeneous extension represents
one of the central aspects of the present analysis. In this case, we derived a
general analytical expression for the shape function in terms of the radial
profile of the polytropic coefficient. Hence, once this function is specified,
the corresponding wormhole geometry is completely determined. This provides a
systematic way of generating and classifying broad families of traversable
wormhole solutions, and makes explicit the direct correspondence between the
radial properties of the supporting matter and the geometry of the resulting
spacetime.

The flare-out condition places important restrictions on the allowed
polytropic parameters. 
By focusing on positive homogeneous and inhomogeneous polytropic
coefficients, we avoid the explicit introduction of phantom or
Chaplygin-type equations of state and find that reality of the
solutions naturally selects odd integer values of the polytropic
exponent. For this
branch the flare-out condition is automatically satisfied, while the necessary
exotic character of the matter  manifests itself
through   the negative-energy sector,
as expected for traversable wormholes in general relativity. Thus, the
polytropic construction provides a qualitatively different realization of
exotic wormhole matter from the more commonly employed phantom or Chaplygin
sources.

In order to investigate explicit configurations, we considered a power-law
radial dependence for the inhomogeneous polytropic coefficient. Although
simple, this ansatz generates a rich variety of geometries depending on the
power-law exponent. We identified qualitatively distinct regimes and obtained
several analytical subclasses, including logarithmic and power-law shape
functions. For a broad part of the parameter space, the non-trivial wormhole
geometry possesses a finite radial support, with the shape function reaching
zero at a finite radius. The matter distribution can therefore be confined to
a bounded region, outside which the interior solution may be replaced by an
appropriate vacuum geometry. We additionally examined a non-integer
polytropic exponent as an illustrative extension of the general construction,
finding an interesting tension between obtaining asymptotic flatness and
maintaining a well-behaved radial polytropic profile.

A particularly interesting result arises for the special relation
$\alpha=2\gamma-3$. In this case, the general solution reduces to a
compact-support configuration of the generalized absurdly benign traversable
wormhole type. The geometry terminates at a finite radius and, for the
constant-redshift solution considered here, can be continued by a flat
exterior region. Hence, a GABTW-like geometry is not imposed from the
beginning, but emerges naturally from a specific realization of the
inhomogeneous polytropic equation of state. This provides a direct connection
between polytropic matter and a class of wormholes designed to localize the
non-trivial geometry and its associated exotic matter.

We further investigated the physical properties and energy conditions of the
obtained solutions. As required by the wormhole flare-out condition, the
radial NEC is necessarily violated at the throat. Nevertheless, the
inhomogeneous polytropic framework introduces additional freedom in
determining how this violation evolves away from the throat, through the radial
dependence of the polytropic coefficient. Consequently, suitable choices of
the model parameters can control the spatial distribution of the exotic
matter. For the GABTW-like branch, the additional analysis presented in the
Appendix shows that the geometry possesses well-defined proper-distance and
embedding properties, while the constant redshift function implies vanishing
traveller acceleration and automatically satisfies the radial tidal-force
constraint. The lateral tidal forces lead instead to a bound on the traversal
velocity. Moreover, the integrated amount of averaged-null-energy-condition
(ANEC)-violating matter remains
finite and, in the vicinity of the throat, can be reduced through an appropriate 
choice of the model parameters.
These properties further support the physical interest of this particular
class of solutions.

The present results show that polytropic matter offers a flexible framework
for constructing traversable wormholes and, importantly, that allowing the
polytropic coefficient to vary radially significantly enlarges the space of
possible geometries. The main novelty of the analysis lies not simply in
obtaining individual wormhole solutions, but in establishing a unified
construction in which homogeneous and inhomogeneous polytropic matter can be
treated systematically, while the latter directly generates different
wormhole geometries through the choice of a single radial function. The
appearance of compact-support and GABTW-like configurations, together with the
possibility of controlling the distribution of the required exotic matter,
illustrates the usefulness of this approach.

There are several directions in which the present analysis can be extended.
An immediate step would be to relax the zero-tidal-force assumption and
investigate non-constant redshift functions, determining whether the additional
freedom can enlarge the physically viable parameter space. It would also be
interesting to study other bounded or asymptotically well-behaved radial
profiles for the inhomogeneous polytropic coefficient, as well as to perform
the explicit junction to exterior Schwarzschild or other vacuum geometries and
analyze the stability of the resulting configurations. Furthermore, one could
extend the construction to rotating or dynamical wormholes, or investigate
polytropic matter in modified-gravity frameworks, where part of the effective
NEC violation may be transferred from the matter sector to the gravitational
one. Finally, a systematic analysis of linear stability and of the response of
these geometries to perturbations would be necessary in order to assess
whether the configurations obtained here can represent not only mathematically
consistent but also dynamically viable traversable wormholes. These 
investigations are left for future projects.

 \begin{acknowledgments}
 E.N. Saridakis acknowledges the contribution of the LISA   CosWG, and of   
COST   
Actions  
 CA21106 ``COSMIC WISPers
in the Dark Universe: Theory, astrophysics and experiments'',  CA21136 
``Addressing observational tensions in cosmology with 
  systematics and fundamental physics (CosmoVerse)'',    CA23130 
``Bridging high and low energies in
search of quantum gravity (BridgeQG)'', and CA24101 ``Testing Fundamental 
Physics with Seismology''.

 \end{acknowledgments}

\appendix{}
\section{Features of the GABTW-like profile}
\label{App}

In this Appendix we investigate in more detail the geometrical and physical
properties of the GABTW-like configuration obtained in
Sec.~\ref{Classes}. In particular, we examine the proper radial distance and
embedding geometry, the gravitational energy, the traversability conditions,
and the total amount of averaged-null-energy-condition
(ANEC)-violating matter. The purpose is to establish
more explicitly the physical characteristics of the compact-support solution
(\ref{GABTWform}) and to compare them with the generalized absurdly benign
wormhole configurations discussed in Ref.~\cite{GABTW}.

\paragraph{Proper radial distance and embedding geometry.}

We begin with the proper radial distance, defined as
\begin{equation}
l(r)=\pm\int_{r_0}^{r}
\frac{\mathrm{d}\tilde r}
{\sqrt{1-\frac{b(\tilde r)}{\tilde r}}}\,,
\label{properdistance}
\end{equation}
where the two signs correspond to the two sides of the wormhole throat.
For the GABTW-like shape function (\ref{GABTWform}), this becomes
\begin{align}
l(r)
=&\pm\int_{r_0}^{r}
\frac{\mathrm{d}\tilde r}
{\sqrt{
1-\frac{r_0}{\tilde r}
\left[1-\mu(\tilde r-r_0)\right]^{\beta}
}}\,,
\nonumber\\
\mu
=&\left(\frac{2n}{2n+1}\right)
\frac{r_0\kappa\rho_0}
{(r_0^2\kappa\rho_0)^{1/(2n+1)}}\,,
\nonumber\\
\beta=&\frac{2n+1}{2n},
\qquad n>1.
\label{l(r)}
\end{align}
The integral cannot in general be expressed in a simple closed form.
Nevertheless, its behavior close to the throat can be obtained analytically.
Expanding around $r=r_0$, we find
\begin{align}
l(r)
\underset{r\rightarrow r_0}{\simeq}
&\pm\frac{1}{\sqrt{1+\beta\mu r_0}}
\int_{r_0}^{r}
\frac{\sqrt{\tilde r}\,\mathrm{d}\tilde r}
{\sqrt{\tilde r-r_0}}
\nonumber\\
=&\pm\frac{1}{\sqrt{1+\beta\mu r_0}}
\bigg[
\sqrt{r}\sqrt{r-r_0}\nonumber\\
& \ \ \ \ \ \ \ \ \ \ \ \  
+r_0
\ln\left(
\sqrt{\frac{r}{r_0}}
+\sqrt{\frac{r}{r_0}-1}
\right)
\bigg] \,,
\label{properapprox}
\end{align}
with
\begin{equation}
r_0\leq r\leq r_0+\frac{1}{\mu} \,.
\end{equation}
Hence, the proper radial distance remains finite throughout the non-trivial
wormhole region.

The proper radial distance is also relevant for estimating the travel time
between two stations located on opposite sides of the throat. Following
Ref.~\cite{Morris:1988cz}, one may choose stations sufficiently far from the
throat such that $1-b(r)/r\simeq1$. Furthermore, since the redshift function
vanishes for the solutions considered here, the coordinate and proper time
intervals coincide.

The spatial geometry can be further characterized through the embedding
function. Considering an equatorial constant-time slice, the embedding
coordinate is determined by
\begin{equation}
z(r)=\pm\int_{r_0}^{r}
\frac{\mathrm{d}\tilde r}
{\sqrt{\frac{\tilde r}{b(\tilde r)}-1}}\,.
\label{embedding}
\end{equation}
For the GABTW-like profile this gives
\begin{equation}
z(r)=\pm\int_{r_0}^{r}
\frac{\mathrm{d}\tilde r}
{\sqrt{
\frac{\tilde r}
{r_0[1-\mu(\tilde r-r_0)]^\beta}
-1
}}\,.
\end{equation}
Expanding again close to the throat, one obtains
\begin{equation}
z(r)
\underset{r\rightarrow r_0}{\simeq}
\pm\frac{\sqrt{r_0}}{\sqrt{1+\beta\mu r_0}}
\int_{r_0}^{r}
\frac{\mathrm{d}\tilde r}
{\sqrt{\tilde r-r_0}}
=
\pm
\frac{2\sqrt{r_0}\sqrt{r-r_0}}
{\sqrt{1+\beta\mu r_0}}\,.
\label{embeddingapprox}
\end{equation}
This behavior explicitly exhibits the standard flare-out geometry in the
vicinity of the throat.

\paragraph{Gravitational energy.}

To further investigate the physical properties of the configuration, we
consider the total gravitational energy associated with a wormhole
\cite{NZCP}, defined as
\begin{equation}
E_G(r)=\!
\int_{r_0}^{r}
\left[
1-\frac{1}{\sqrt{1\!-\!b(\tilde r)/\tilde r}}
\right]
\rho(\tilde r)\tilde r^2\,\mathrm{d}\tilde r
+\frac{r_0}{2G}
=M-M^P,
\label{Eg}
\end{equation}
where $M$ and $M^P$ denote the total and proper masses, respectively.

Since the non-trivial matter distribution has compact support,
$r_0\leq r\leq r_0+1/\mu$, it is natural to evaluate these quantities over
this interval. The total mass is
\begin{equation}
M=
\int_{r_0}^{r_0+\frac{1}{\mu}}
4\pi\rho(\tilde r)\tilde r^2\,\mathrm{d}\tilde r
=
-\frac{4\pi}{\kappa}r_0\,,
\label{Mtotal}
\end{equation}
while the proper mass is
\begin{align}
M^P
=&\pm
\int_{r_0}^{r_0+\frac{1}{\mu}}
\frac{
4\pi\rho(\tilde r)\tilde r^2
}{
\sqrt{1-b(\tilde r)/\tilde r}
}
\,\mathrm{d}\tilde r
\nonumber\\
=&\pm\frac{4\pi}{\kappa}
\int_{r_0}^{r_0+\frac{1}{\mu}}
\frac{
b'(\tilde r)
}{
\sqrt{1-b(\tilde r)/\tilde r}
}
\,\mathrm{d}\tilde r
\nonumber\\
\simeq&
\pm\frac{4\pi}{\kappa}
\left[
-\frac{
2\beta r_0\sqrt{\mu r_0}
}{
\sqrt{1+\beta\mu r_0}
}
\right].
\label{Mp}
\end{align}
The sign again distinguishes the two sides of the wormhole. Consequently,
the gravitational energy becomes
\begin{equation}
E_G
\simeq
-\frac{4\pi}{\kappa}r_0
\left[
1\mp
\frac{
2\beta\sqrt{\mu r_0}
}{
\sqrt{1+\beta\mu r_0}
}
\right].
\label{EGapprox}
\end{equation}
In the regime $\mu r_0\gg1$, the latter approaches
\begin{equation}
E_G
\simeq
-\frac{4\pi}{\kappa}r_0
\left(
1\mp2\sqrt{\beta}
\right),
\label{EGlarge}
\end{equation}
showing that the gravitational-energy contribution remains finite.

\paragraph{Traversability conditions.}

An important requirement for a traversable wormhole is that the proper
acceleration experienced by the traveller remains below a tolerable value,
which is conventionally taken to be the Earth's gravitational acceleration,
$g_\oplus\simeq980\,{\rm cm/s^2}$. In the traveller's orthonormal frame, the
corresponding condition reads
\begin{equation}
\left|
\sqrt{1-\frac{b(r)}{r}}\,
e^{-\Phi(r)}
\left(
\Gamma e^{\Phi(r)}
\right)'
\right|
\leq
\frac{g_\oplus}{c^2}\,,
\label{acceleration}
\end{equation}
where $\Gamma$ denotes the Lorentz factor of the traveller. Since
$\Phi(r)=0$ in the present construction, a traveller moving with constant
velocity experiences vanishing proper acceleration, consistently with the
standard Morris-Thorne analysis \cite{Morris:1988cz}.

The tidal accelerations must also remain sufficiently small. The radial
tidal-force constraint is
\begin{eqnarray}
&&
\!\!\!\!\!\!\!\!\!\!\!\!\!\!\!\!
\left|
\left[1-\frac{b(r)}{r}\right]
\left[
\Phi''(r)+\Phi'^2(r)
-\frac{b'(r)r-b(r)}
{2r[r-b(r)]}\Phi'(r)
\right]
\right|\nonumber\\
&&
\ \ 
\cdot
c^2
\left|\eta^{\hat 1'}\right|
\leq g_\oplus \,.
\label{RTC}
\end{eqnarray}
For $\Phi(r)=0$, this condition is identically satisfied.

The lateral tidal-force constraint reads
{\small{
\begin{equation}
\left|
\frac{\Gamma^2c^2}{2r^2}
\left[
\frac{v^2(r)}{c^2}
\left(
b'(r)-\frac{b(r)}{r}
\right)
+
2r[r-b(r)]\Phi'(r)
\right]
\right|
|\eta|
\leq g_\oplus \,.
\label{LTC}
\end{equation}}}
Here $\eta$ represents the characteristic size of the traveller. Following
Ref.~\cite{Morris:1988cz}, we take $|\eta|\simeq2$. Assuming approximately 
constant velocity and
$\Gamma\simeq1$ close to the throat, the lateral constraint reduces to
\begin{equation}
\frac{v^2(r_0)}{r_0^2}
\left(
r_0\beta\mu+1
\right)
\lesssim g_\oplus \,,
\label{LTCt}
\end{equation}
which yields the bound
\begin{equation}
v
\lesssim
r_0
\sqrt{
\frac{g_\oplus}
{r_0\beta\mu+1}
}\,.
\label{vbound}
\end{equation}
Thus, the lateral tidal force constrains the allowed traversal velocity.
For vanishing velocity the corresponding tidal contribution vanishes.

Using this estimate, the characteristic crossing time may be written as
\begin{equation}
\Delta t
\simeq
2\times10^4 \, \frac{r_0}{v}
\simeq
2\times10^4
\sqrt{
\frac{r_0\beta\mu+1}{g_\oplus}
}\,.
\label{crossingtime}
\end{equation}
Hence, the crossing time depends explicitly on the throat scale and on the
parameter $\mu$ that controls the radial extent of the GABTW-like region.

\paragraph{Amount of ANEC-violating matter.}

Finally, we quantify the total amount of 
ANEC-violating matter using the volume-integral quantifier introduced in
Ref.~\cite{VKD}. For the metric (\ref{GABTWform}), one has
\begin{align}
I_V
=&
\frac{1}{\kappa}
\int_{r_0}^{r_0+\frac{1}{\mu}}
[r-b(r)]
\left[
\ln\left(
\frac{e^{2\Phi(r)}}{1-b(r)/r}
\right)
\right]'
\mathrm{d}r
\nonumber\\
=&
\frac{1}{\kappa}
\int_{r_0}^{r_0+\frac{1}{\mu}}
\left[
\frac{b(r)}{r}-b'(r)
\right]
\mathrm{d}r
\label{IV}\\
\simeq&
\frac{1}{\kappa}
\int_{r_0}^{r_0+\frac{1}{\mu}}
\left[
\beta\mu r_0+1
+c(r)(r-r_0)
\right]
\mathrm{d}r\,,
\nonumber
\end{align}
where in the last step we have expanded the integrand in the vicinity of the
throat. Introducing the corresponding coefficient $c(r)$, the integration
gives
\begin{equation}
I_V
=
\frac{1}{\kappa}
\left[
\frac{3}{2}\beta r_0
-\frac{1}{2}\beta^2r_0
-\frac{\beta}{2\mu}
+\frac{1}{\mu}
-\frac{1}{2\mu^2r_0}
\right].
\label{IVresult}
\end{equation}
For $\mu r_0\gg1$, this reduces to
\begin{equation}
I_V
\simeq
\frac{1}{\kappa}
\left(
\frac{3}{2}\beta r_0
-\frac{1}{2}\beta^2r_0
\right),
\label{IVlarge}
\end{equation}
which remains finite. Therefore, the ANEC-violating matter associated with
the GABTW-like configuration is finite and, through an appropriate choice of
the parameters controlling the near-throat region, its integrated amount can
be reduced. This complements the local energy-condition analysis of
Sec.~\ref{Classes} and further illustrates the physical interest of the
compact-support solution.


\begin{thebibliography}{9999}      

 

 




\bibitem{Morris:1988cz}
M.~S.~Morris and K.~S.~Thorne,
Am. J. Phys. \textbf{56}, 395-412 (1988).




   
   

\bibitem {Visser:1995cc} M. Visser, \textit{Lorentzian Wormholes: From Einstein 
to
Hawking}, AIP Press, New York, (1995).



 

\bibitem{Morris:1988tu}
M.~S.~Morris, K.~S.~Thorne and U.~Yurtsever,
Phys. Rev. Lett. \textbf{61}, 1446-1449 (1988).



\bibitem{Visser:1989kh}
M.~Visser,
Phys. Rev. D \textbf{39}, 3182-3184 (1989)
   [\href{https://arxiv.org/abs/0809.0907}{arXiv:0809.0907 [gr-qc]}].


   
  

\bibitem{Roman:1992xj}
T.~A.~Roman,
Phys. Rev. D \textbf{47}, 1370-1379 (1993)
   [\href{https://arxiv.org/abs/gr-qc/9211012}{arXiv:gr-qc/9211012 [gr-qc]}].

\bibitem{Armendariz-Picon:2002km}
C.~Armendariz-Picon,
Phys. Rev. D \textbf{65}, 104010 (2002)
   [\href{https://arxiv.org/abs/gr-qc/0201027}{arXiv:gr-qc/0201027 [gr-qc]}].


\bibitem{Lemos:2003jb}
J.~P.~S.~Lemos, F.~S.~N.~Lobo and S.~Quinet de Oliveira,
Phys. Rev. D \textbf{68}, 064004 (2003)
   [\href{https://arxiv.org/abs/gr-qc/0302049}{arXiv:gr-qc/0302049 [gr-qc]}].

\bibitem{Kar:2004hc}
S.~Kar, N.~Dadhich and M.~Visser,
Pramana \textbf{63}, 859-864 (2004)
   [\href{https://arxiv.org/abs/gr-qc/0405103}{arXiv:gr-qc/0405103 [gr-qc]}].

\bibitem{Sushkov:2005kj}
S.~V.~Sushkov,
Phys. Rev. D \textbf{71}, 043520 (2005)
   [\href{https://arxiv.org/abs/gr-qc/0502084}{arXiv:gr-qc/0502084 [gr-qc]}].

\bibitem{Lobo:2005us}
F.~S.~N.~Lobo,
Phys. Rev. D \textbf{71}, 084011 (2005)
   [\href{https://arxiv.org/abs/gr-qc/0502099}{arXiv:gr-qc/0502099 [gr-qc]}].

\bibitem{Lobo:2005yv}
F.~S.~N.~Lobo,
Phys. Rev. D \textbf{71}, 124022 (2005)
   [\href{https://arxiv.org/abs/gr-qc/0506001}{arXiv:gr-qc/0506001 [gr-qc]}].

\bibitem{Zaslavskii:2005fs}
O.~B.~Zaslavskii,
Phys. Rev. D \textbf{72}, 061303 (2005)
   [\href{https://arxiv.org/abs/gr-qc/0508057}{arXiv:gr-qc/0508057 [gr-qc]}].

\bibitem{Lobo:2005vc}
F.~S.~N.~Lobo,
Phys. Rev. D \textbf{73}, 064028 (2006)
   [\href{https://arxiv.org/abs/gr-qc/0511003}{arXiv:gr-qc/0511003 [gr-qc]}].

\bibitem{Rahaman:2005ur}
F.~Rahaman, M.~Kalam, M.~Sarker and K.~Gayen,
Phys. Lett. B \textbf{633}, 161-163 (2006)
   [\href{https://arxiv.org/abs/gr-qc/0512075}{arXiv:gr-qc/0512075 [gr-qc]}].

\bibitem{Kuhfittig:2006rd}
P.~K.~F.~Kuhfittig,
Class. Quant. Grav. \textbf{23}, 5853-5860 (2006)
   [\href{https://arxiv.org/abs/gr-qc/0608055}{arXiv:gr-qc/0608055 [gr-qc]}].


\bibitem{Chakraborty:2007na}
S.~Chakraborty and T.~Bandyopadhyay,
Int. J. Mod. Phys. D \textbf{18}, 463-476 (2009)
   [\href{https://arxiv.org/abs/0707.1183}{arXiv:0707.1183 [gr-qc]}].


\bibitem{Gonzalez:2009hn}
J.~A.~Gonzalez, F.~S.~Guzman, N.~Montelongo-Garcia and T.~Zannias,
Phys. Rev. D \textbf{79}, 064027 (2009)
   [\href{https://arxiv.org/abs/0906.5590}{arXiv:0906.5590 [gr-qc]}].

\bibitem{Gonzalez:2008wd}
J.~A.~Gonzalez, F.~S.~Guzman and O.~Sarbach,
Class. Quant. Grav. \textbf{26}, 015010 (2009)
   [\href{https://arxiv.org/abs/0806.0608}{arXiv:0806.0608 [gr-qc]}].

\bibitem{Gonzalez:2008xk}
J.~A.~Gonzalez, F.~S.~Guzman and O.~Sarbach,
Class. Quant. Grav. \textbf{26}, 015011 (2009)
   [\href{https://arxiv.org/abs/0806.1370}{arXiv:0806.1370 [gr-qc]}].



\bibitem{Parsaei:2019hji}
F.~Parsaei and S.~Rastgoo,
Phys. Rev. D \textbf{99}, 104037 (2019)
   [\href{https://arxiv.org/abs/1903.08251}{arXiv:1903.08251 [gr-qc]}].



\bibitem{Garattini:2019ivd}
R.~Garattini,
Eur. Phys. J. C \textbf{79}, 951 (2019)
   [\href{https://arxiv.org/abs/1907.03623}{arXiv:1907.03623 [gr-qc]}].


\bibitem{Parsaei:2019utg}
F.~Parsaei and S.~Rastgoo,
Eur. Phys. J. C \textbf{80}, no.5, 366 (2020)
   [\href{https://arxiv.org/abs/1909.09899}{arXiv:1909.09899 [gr-qc]}].


\bibitem{Alfaro:2024tdr}
S.~Alfaro, P.~A.~Gonz{\'a}lez, D.~Olmos, E.~Papantonopoulos and Y.~V{\'a}squez,
Phys. Rev. D \textbf{109}, no.10, 104009 (2024)
   [\href{https://arxiv.org/abs/2402.15575}{arXiv:2402.15575 [gr-qc]}].

\bibitem{CANTATA:2021asi}
E.~N.~Saridakis \textit{et al.} [CANTATA],
Springer, (2021),
   [\href{https://arxiv.org/abs/2105.12582}{arXiv:2105.12582 [gr-qc]}].
   
 
 



 
\bibitem{Lobo:2017cay}
F.~S.~N.~Lobo,
Fundam. Theor. Phys. \textbf{189}, 279 (2017)
Springer, (2017), 
 [\href{https://arxiv.org/abs/2103.05610}{arXiv:2103.05610 [gr-qc]}].
   
 
  




\bibitem{Lobo:2009ip}
F.~S.~N.~Lobo and M.~A.~Oliveira,
Phys. Rev. D \textbf{80}, 104012 (2009)
   [\href{https://arxiv.org/abs/0909.5539}{arXiv:0909.5539 [gr-qc]}].

\bibitem{GarciaLobo:2010}
N.~M.~Garcia and F.~S.~N.~Lobo,
Phys. Rev. D \textbf{82}, 104018 (2010)
   [\href{https://arxiv.org/abs/1007.3040}{arXiv:1007.3040 [gr-qc]}].

\bibitem{GarciaLobo:2011}
N.~M.~Garcia and F.~S.~N.~Lobo,
Class. Quant. Grav. \textbf{28}, 085018 (2011)
   [\href{https://arxiv.org/abs/1012.2443}{arXiv:1012.2443 [gr-qc]}].

\bibitem{Lobo:2008zu}
F.~S.~N.~Lobo,
Class. Quant. Grav. \textbf{25}, 175006 (2008)
   [\href{https://arxiv.org/abs/0801.4401}{arXiv:0801.4401 [gr-qc]}].

\bibitem{Lobo:2007qi}
F.~S.~N.~Lobo,
Phys. Rev. D \textbf{75}, 064027 (2007)
   [\href{https://arxiv.org/abs/gr-qc/0701133}{arXiv:gr-qc/0701133 [gr-qc]}].

\bibitem{Harko:2013aya}
T.~Harko, F.~S.~N.~Lobo, M.~K.~Mak and S.~V.~Sushkov,
Phys. Rev. D \textbf{87}, 067504 (2013)
   [\href{https://arxiv.org/abs/1301.6878}{arXiv:1301.6878 [gr-qc]}].

\bibitem{SharifNawazish:2018}
M.~Sharif and I.~Nawazish,
Annals Phys. \textbf{389}, 283-305 (2018)
   [\href{https://arxiv.org/abs/1801.05022}{arXiv:1801.05022 [gr-qc]}].

\bibitem{SamantaGodaniBamba:2020}
G.~C.~Samanta, N.~Godani and K.~Bamba,
Int. J. Mod. Phys. D \textbf{29}, 2050068 (2020)
   [\href{https://arxiv.org/abs/1811.06834}{arXiv:1811.06834 [gr-qc]}].

\bibitem{GhoshMitra:2021}
B.~Ghosh and S.~Mitra,
Int. J. Mod. Phys. A \textbf{36}, 2150119 (2021)
   [\href{https://arxiv.org/abs/2108.12670}{arXiv:2108.12670 [gr-qc]}].

\bibitem{AgrawalMishra:2022}
A.~S.~Agrawal, B.~Mishra, F.~Tello-Ortiz and A.~Alvarez,
Fortsch. Phys. \textbf{70}, 2100177 (2022)
   [\href{https://arxiv.org/abs/2112.01013}{arXiv:2112.01013 [gr-qc]}].

\bibitem{BaruahGoswami:2022}
A.~Baruah, P.~Goswami and A.~Deshamukhya,
Int. J. Mod. Phys. D \textbf{31}, 2250119 (2022)
   [\href{https://arxiv.org/abs/2111.02941}{arXiv:2111.02941 [gr-qc]}].







\bibitem{ZubairWaheedAhmad:2016}
M.~Zubair, S.~Waheed and Y.~Ahmad,
Eur. Phys. J. C \textbf{76}, 444 (2016)
   [\href{https://arxiv.org/abs/1607.05998}{arXiv:1607.05998 [gr-qc]}].


\bibitem{MoraesSahoo:2018a}
P.~H.~R.~S.~Moraes and P.~K.~Sahoo,
Phys. Rev. D \textbf{97}, 024007 (2018)
   [\href{https://arxiv.org/abs/1709.00027}{arXiv:1709.00027 [gr-qc]}].

\bibitem{SahooMoraesSahoo:2018}
P.~K.~Sahoo, P.~H.~R.~S.~Moraes and P.~Sahoo,
Eur. Phys. J. C \textbf{78}, 46 (2018)
   [\href{https://arxiv.org/abs/1709.07774}{arXiv:1709.07774 [gr-qc]}].

\bibitem{MoraesDePaulaCorrea:2019}
P.~H.~R.~S.~Moraes, W.~de Paula and R.~A.~C.~Correa,
Int. J. Mod. Phys. D \textbf{28}, 1950098 (2019)
   [\href{https://arxiv.org/abs/1710.07680}{arXiv:1710.07680 [gr-qc]}].

\bibitem{MoraesSahoo:2019}
P.~H.~R.~S.~Moraes and P.~K.~Sahoo,
Eur. Phys. J. C \textbf{79}, 677 (2019)
   [\href{https://arxiv.org/abs/1903.03421}{arXiv:1903.03421 [gr-qc]}].

\bibitem{BanerjeeJasimGhosh:2021}
A.~Banerjee, M.~K.~Jasim and S.~G.~Ghosh,
Annals Phys. \textbf{433}, 168575 (2021)
   [\href{https://arxiv.org/abs/2003.01545}{arXiv:2003.01545 [gr-qc]}].

\bibitem{SahooMoraesLapola:2021}
P.~Sahoo, P.~H.~R.~S.~Moraes, M.~M.~Lapola and P.~K.~Sahoo,
Int. J. Mod. Phys. D \textbf{30}, 2150100 (2021)
   [\href{https://arxiv.org/abs/2012.00258}{arXiv:2012.00258 [gr-qc]}].


\bibitem{SaleemAslam:2023}
R.~Saleem and M.~I.~Aslam,
Chin. J. Phys. \textbf{85}, 741-751 (2023)
.

\bibitem{Nashed:2026wdv}
G.~G.~L.~Nashed, W.~El Hanafy, A.~Abebe, K.~Bamba and E.~N.~Saridakis,
   [\href{https://arxiv.org/abs/2606.01141}{arXiv:2606.01141 [gr-qc]}].


\bibitem{Sharif:2013exa}
M.~Sharif and S.~Rani,
Phys. Rev. D \textbf{88}, no.12, 123501 (2013).


\bibitem{Kofinas:2015hla}
G.~Kofinas, E.~Papantonopoulos and E.~N.~Saridakis,
Phys. Rev. D \textbf{91}, no.10, 104034 (2015)
   [\href{https://arxiv.org/abs/1501.00365}{arXiv:1501.00365 [gr-qc]}].


\bibitem{MehdizadehZiaie:2017}
M.~R.~Mehdizadeh and A.~H.~Ziaie,
Phys. Rev. D \textbf{95}, 064049 (2017)
   [\href{https://arxiv.org/abs/1704.06923}{arXiv:1704.06923 [gr-qc]}].


\bibitem{Mehdizadeh:2017dhb}
M.~R.~Mehdizadeh and A.~H.~Ziaie,
Phys. Rev. D \textbf{96}, no.12, 124017 (2017)
   [\href{https://arxiv.org/abs/1709.09028}{arXiv:1709.09028 [gr-qc]}].


\bibitem{SaaidiTavakoli:2021}
K.~Saaidi and S.~Tavakoli,
Phys. Dark Univ. \textbf{31}, 100763 (2021)
.
\bibitem{Landry:2025whg}
A.~Landry, Y.~Sekhmani, S.~K.~Maurya, A.~Ali and E.~N.~Saridakis,
   [\href{https://arxiv.org/abs/2508.06290}{arXiv:2508.06290 [gr-qc]}].



\bibitem{ParsaeiRastgooSahoo:2022}
F.~Parsaei, S.~Rastgoo and P.~K.~Sahoo,
Eur. Phys. J. Plus \textbf{137}, 1083 (2022)
   [\href{https://arxiv.org/abs/2203.06374}{arXiv:2203.06374 [gr-qc]}].

\bibitem{MustafaHassanSahoo:2022}
G.~Mustafa, Z.~Hassan and P.~K.~Sahoo,
Annals Phys. \textbf{437}, 168751 (2022)
   [\href{https://arxiv.org/abs/2112.15112}{arXiv:2112.15112 [gr-qc]}].

\bibitem{SokoliukHassanSahoo:2022}
O.~Sokoliuk, Z.~Hassan, P.~K.~Sahoo and A.~Baransky,
Annals Phys. \textbf{443}, 168968 (2022)
   [\href{https://arxiv.org/abs/2201.00743}{arXiv:2201.00743 [gr-qc]}].

\bibitem{KiroriwalKumarMaurya:2023}
S.~Kiroriwal, J.~Kumar, S.~K.~Maurya and S.~Chaudhary,
Phys. Scripta \textbf{98}, 125305 (2023)
.

\bibitem{RastgooParsaei:2024}
S.~Rastgoo and F.~Parsaei,
Eur. Phys. J. C \textbf{84}, 563 (2024)
   [\href{https://arxiv.org/abs/2402.15178}{arXiv:2402.15178 [gr-qc]}].

\bibitem{HohmannKaranasou:2025}
M.~Hohmann and V.~Karanasou,
Phys. Rev. D \textbf{111}, 064057 (2025)
   [\href{https://arxiv.org/abs/2412.11730}{arXiv:2412.11730 [gr-qc]}].


\bibitem{DehghaniDayyani:2009}
M.~H.~Dehghani and Z.~Dayyani,
Phys. Rev. D \textbf{79}, 064010 (2009)
   [\href{https://arxiv.org/abs/0903.4262}{arXiv:0903.4262 [gr-qc]}].

\bibitem{Kanti:2011jz}
P.~Kanti, B.~Kleihaus and J.~Kunz,
Phys. Rev. D \textbf{85}, 044007 (2012)
   [\href{https://arxiv.org/abs/1111.4049}{arXiv:1111.4049 [hep-th]}].

\bibitem{MehdizadehZangenehLobo:2015}
M.~R.~Mehdizadeh, M.~Kord Zangeneh and F.~S.~N.~Lobo,
Phys. Rev. D \textbf{92}, 044022 (2015)
   [\href{https://arxiv.org/abs/1506.03427}{arXiv:1506.03427 [gr-qc]}].

\bibitem{ZangenehLoboDehghani:2015}
M.~Kord Zangeneh, F.~S.~N.~Lobo and M.~H.~Dehghani,
Phys. Rev. D \textbf{92}, 124049 (2015)
   [\href{https://arxiv.org/abs/1510.07089}{arXiv:1510.07089 [gr-qc]}].

\bibitem{MehdizadehZiaie:2021}
M.~R.~Mehdizadeh and A.~H.~Ziaie,
Phys. Rev. D \textbf{104}, 104050 (2021)
   [\href{https://arxiv.org/abs/2111.14828}{arXiv:2111.14828 [gr-qc]}].

\bibitem{ChakrabortyChakraborty:2025}
M.~Chakraborty and S.~Chakraborty,
Phys. Dark Univ. \textbf{47}, 101793 (2025)
.

\bibitem{MunizEtAl:2025}
C.~R.~Muniz, M.~B.~Cruz, R.~M.~P.~Neves, M.~Farooq and M.~Zubair,
JCAP \textbf{07}, 015 (2025)
   [\href{https://arxiv.org/abs/2503.12943}{arXiv:2503.12943 [hep-th]}].


\bibitem{Tsilioukas:2023tdw}
S.~A.~Tsilioukas, E.~N.~Saridakis and C.~Tzerefos,
Phys. Rev. D \textbf{109}, no.8, 084010 (2024)
   [\href{https://arxiv.org/abs/2312.07486}{arXiv:2312.07486 [gr-qc]}].

\bibitem{Capozziello:2018mqy}
S.~Capozziello, R.~Pincak and E.~N.~Saridakis,
Annals Phys. \textbf{390}, 303-333 (2018).

\bibitem{Chatzifotis:2022mob}
N.~Chatzifotis, P.~Dorlis, N.~E.~Mavromatos and E.~Papantonopoulos,
Phys. Rev. D \textbf{105}, no.8, 084051 (2022)
   [\href{https://arxiv.org/abs/2202.03496}{arXiv:2202.03496 [gr-qc]}].



\bibitem{Barcelo:2000zf} 
C.~Barcelo and M.~Visser,
Class. Quant. Grav. \textbf{17}, 3843-3864 (2000)
   [\href{https://arxiv.org/abs/gr-qc/0003025}{arXiv:gr-qc/0003025 [gr-qc]}].

\bibitem{KorolevSushkov:2014}
R.~V.~Korolev and S.~V.~Sushkov,
Phys. Rev. D \textbf{90}, 124025 (2014)
   [\href{https://arxiv.org/abs/1408.1235}{arXiv:1408.1235 [gr-qc]}].

\bibitem{BakopoulosCharmousisKanti:2022}
A.~Bakopoulos, C.~Charmousis and P.~Kanti,
JCAP \textbf{05}, 022 (2022)
   [\href{https://arxiv.org/abs/2111.09857}{arXiv:2111.09857 [gr-qc]}].

\bibitem{BakopoulosChatzifotis:2023}
A.~Bakopoulos, N.~Chatzifotis, C.~Erices and E.~Papantonopoulos,
JCAP \textbf{11}, 055 (2023)
   [\href{https://arxiv.org/abs/2306.16768}{arXiv:2306.16768 [hep-th]}].

\bibitem{Chatzifotis:2021hpg}
N.~Chatzifotis, E.~Papantonopoulos and C.~Vlachos,
Phys. Rev. D \textbf{105}, no.6, 064025 (2022)
   [\href{https://arxiv.org/abs/2111.08773}{arXiv:2111.08773 [gr-qc]}].


\bibitem{CapozzielloHarkoKoivisto:2012}
S.~Capozziello, T.~Harko, T.~S.~Koivisto, F.~S.~N.~Lobo and G.~J.~Olmo,
Phys. Rev. D \textbf{86}, 127504 (2012)
   [\href{https://arxiv.org/abs/1209.5862}{arXiv:1209.5862 [gr-qc]}].

\bibitem{BambiCardenasOlmo:2016}
C.~Bambi, A.~Cardenas-Avendano, G.~J.~Olmo and D.~Rubiera-Garcia,
Phys. Rev. D \textbf{93}, 064016 (2016)
   [\href{https://arxiv.org/abs/1511.03755}{arXiv:1511.03755 [gr-qc]}].

\bibitem{RosaLemosLobo:2018}
J.~L.~Rosa, J.~P.~S.~Lemos and F.~S.~N.~Lobo,
Phys. Rev. D \textbf{98}, 064054 (2018)
   [\href{https://arxiv.org/abs/1808.08975}{arXiv:1808.08975 [gr-qc]}].

\bibitem{LoboOlmoOrazi:2020}
F.~S.~N.~Lobo, G.~J.~Olmo, E.~Orazi, D.~Rubiera-Garcia and A.~Rustam,
Phys. Rev. D \textbf{102}, 104012 (2020)
   [\href{https://arxiv.org/abs/2009.10997}{arXiv:2009.10997 [gr-qc]}].

\bibitem{KordZangenehLobo:2021}
M.~Kord Zangeneh and F.~S.~N.~Lobo,
Eur. Phys. J. C \textbf{81}, 285 (2021)
   [\href{https://arxiv.org/abs/2011.01745}{arXiv:2011.01745 [gr-qc]}].

\bibitem{Rosa:2021}
J.~L.~Rosa,
Phys. Rev. D \textbf{104}, 064002 (2021)
   [\href{https://arxiv.org/abs/2107.14225}{arXiv:2107.14225 [gr-qc]}].
   
   
   

\bibitem{HarkoLoboMakSushkov:2015}
T.~Harko, F.~S.~N.~Lobo, M.~K.~Mak and S.~V.~Sushkov,
Mod. Phys. Lett. A \textbf{30}, 1550190 (2015)
   [\href{https://arxiv.org/abs/1307.1883}{arXiv:1307.1883 [gr-qc]}].

\bibitem{Shaikh:2015}
R.~Shaikh,
Phys. Rev. D \textbf{92}, 024015 (2015)
   [\href{https://arxiv.org/abs/1505.01314}{arXiv:1505.01314 [gr-qc]}].



\bibitem{NandiIslamEvans:1997}
K.~K.~Nandi, A.~Islam and J.~Evans,
Phys. Rev. D \textbf{55}, 2497-2500 (1997)
   [\href{https://arxiv.org/abs/0906.0436}{arXiv:0906.0436 [gr-qc]}].


\bibitem{ZiaieMehdizadeh:2024}
A.~H.~Ziaie and M.~R.~Mehdizadeh,
Class. Quant. Grav. \textbf{41}, 145001 (2024)
   [\href{https://arxiv.org/abs/2406.10821}{arXiv:2406.10821 [gr-qc]}].


\bibitem{Papantonopoulos:2019ugr}
E.~Papantonopoulos and C.~Vlachos,
Phys. Rev. D \textbf{101}, no.6, 064025 (2020)
   [\href{https://arxiv.org/abs/1912.04005}{arXiv:1912.04005 [gr-qc]}].




 




\bibitem{Tooper1964}
R.~F.~Tooper,
Astrophys.\ J.\ \textbf{140} (1964) 434.

\bibitem{Tooper1965}
R.~F.~Tooper,
Astrophys.\ J.\ \textbf{142} (1965) 1541.

\bibitem{Cataldo2013}
M.~Cataldo, F.~Ar{\'o}stica and S.~Bahamonde,
Eur. Phys. J. C \textbf{73} (2013) 2517
   [\href{https://arxiv.org/abs/1307.4122}{arXiv:1307.4122 [gr-qc]}].



\bibitem{VKD}
M.~Visser, S.~Kar and N.~Dadhich,
Phys. Rev. Lett. \textbf{90}, 201102 (2003)
   [\href{https://arxiv.org/abs/gr-qc/0301003}{arXiv:gr-qc/0301003 [gr-qc]}].


\bibitem{Parsaeia2020}
F.~Parsaei and S.~Rastgoo,
Eur. Phys. J. C \textbf{80} (2020) 366
   [\href{https://arxiv.org/abs/1909.09899}{arXiv:1909.09899 [gr-qc]}].

\bibitem{Jamil2009}
M.~Jamil, P.~K.~F.~Kuhfittig, F.~Rahaman and S.~A.~Rakib,
Eur. Phys. J. C \textbf{67} (2010)  513-520
   [\href{https://arxiv.org/abs/0906.2142}{arXiv:0906.2142 [gr-qc]}].

\bibitem{ThinShell2024}
G.~Mustafa, F.~Javed, S.~K.~Maurya and S.~Ray,
Chin. J. Phys. \textbf{88} (2024)  32
   [\href{https://arxiv.org/abs/2211.10778}{arXiv:2211.10778 [gr-qc]}].



 
\bibitem{Lobo:2006mt} 
F.~S.~N.~Lobo,
   [\href{https://arxiv.org/abs/gr-qc/0611150}{arXiv:gr-qc/0611150 [gr-qc]}].

\bibitem{g0}
J.~Binney and S.~Tremaine,
\textit{Galactic Dynamics}, Princeton University Press, Princeton, (2008).


\bibitem{g1}
S.~Chandrasekhar,
Rev.\ Mod.\ Phys.\ \textbf{15} (1943) 1.

\bibitem{g2}
S.~Chandrasekhar,
\textit{An Introduction to the Study of Stellar Structure}, University of 
Chicago Press, Chicago, (1939).



\bibitem{g3}
M.~K.~Mak and T.~Harko,
Proc.\ Roy.\ Soc.\ Lond.\ A \textbf{459} (2003) 393.




\bibitem{g4}
O.~Bertolami, A.~A.~Sen, S.~Sen, and P.~T.~Silva,
Phys.\ Rev.\ D \textbf{70} (2004) 083506
   [\href{https://arxiv.org/abs/astro-ph/0402387}{arXiv:astro-ph/0402387}].

\bibitem{g5}
M.~C.~Bento, O.~Bertolami, and A.~A.~Sen,
Phys.\ Rev.\ D \textbf{66} (2002) 043507
   [\href{https://arxiv.org/abs/gr-qc/0202064}{arXiv:gr-qc/0202064}].



\bibitem {Isr} W.~Israel, Il Nuovo Cimento B Series \textbf{10 44.1} (1966) 
1-14.


\bibitem{GABTW}
R.~Garattini,
Eur. Phys. J. C \textbf{80}, no.12, 1172 (2020)
   [\href{https://arxiv.org/abs/2008.05901}{arXiv:2008.05901 [gr-qc]}].


\bibitem{NZCP} 
K.~K.~Nandi, Y.~Z.~Zhang, R.~G.~Cai and A.~Panchenko,
Phys. Rev. D \textbf{79}, 024011 (2009)
   [\href{https://arxiv.org/abs/0809.4143}{arXiv:0809.4143 [gr-qc]}].



\end{thebibliography}
\end{document}